\documentclass[amssymb,aps,floatfix,showkeys,showpacs,twocolumn]{revtex4}

\usepackage{graphicx}
\usepackage{amsmath}

\begin{document}

\title[Clebsch--Gordan Series of Quaternions]{Clebsch--Gordan Coefficients of the Quaternion Group}

\author{Richard J. Mathar}
\homepage{http://www.strw.leidenuniv.nl/~mathar}
\email{mathar@strw.leidenuniv.nl}
\affiliation{Leiden Observatory, P.O. Box 9513, 2300 RA Leiden, The Netherlands}


\date{\today}

\begin{abstract}
The Clebsch--Gordan coefficients of the Kronecker products of 
the irreducible representations of the Quaternion Group $Q_8$,
of the Generalized Quaternion Groups $Q_{16}$ and $Q_{32}$, and of the
factor group $(Q_{32}\times Q_{32})/\{(1,1),(-1,-1)\}$ are computed as
eigenvectors of a well-known matrix of
triple-products of
the irreducible representations.
\end{abstract}

\maketitle
\section{Scope} 

The quaternion group $Q_8$ is a non-abelian group of order 8 \cite{MillerPAPS37,CoxeterCPAM26,GirardEJP5}.
In relativistic quantum-mechanics, the quaternion algebra appears as a representation
of Dirac bi-spinors \cite{SaueJCP111}; the direct product
$Q_8\times Q_8$ spans the algebra of the 2-fermion matrix elements \cite{VisscherJCC23}.

Whereas the Clebsch--Gordan (CG) decomposition of other small groups is
well documented serving studies of point group operations
in molecular and solid-state physics \cite{Flurry,Koster},
the CG series of $Q_8$---although simple---seems
not to be readily available. The manuscript follows standard concepts to
reduce the product of irreducible representations of some
quaternionic groups of low order \cite{StringhamAJM4}.

\section{Structure of $Q_8$} 
\subsection{Multiplication Table, Classes}
$Q_8$ is the fourth group of order 8 in the
GAP4 library \cite{BescheIJAC12,BescheJSC27,GAP4,FieldsteelASP2008}, the fifth
in the Schaps enumeration \cite{Schaps}.
It is called $\Gamma_2a_2$ by Hall--Senior \cite{Hall},
$8/5$ in the Thomas-Wood enumeration \cite{Thomas},
and apparently this index carries over as $8.5$ \cite{HalesJA316}.
The $|Q_8|=8$ elements $g_i$ are enumerated with their standard
symbols in Table \ref{tab.namq8}.

\begin{table}
\caption{Indices of the $Q_8$ group elements $g_j$, their standard names, and orders.}
\begin{ruledtabular}
\begin{tabular}{r|rrrrrrrrrrrrrrrr}
$j$ & 1 & 2 & 3 & 4 & 5 & 6 & 7 & 8 \\
\hline
$g_j$ & \texttt{1} & \texttt{-1} & \texttt{i} & \texttt{-i} & \texttt{j} & \texttt{-j} & \texttt{k} & \texttt{-k} \\
 & 1 & 2 & 4 & 4 & 4 & 4 & 4 & 4 \\
\end{tabular}
\end{ruledtabular}
\label{tab.namq8}
\end{table}

\begin{figure}
\includegraphics[scale=0.3]{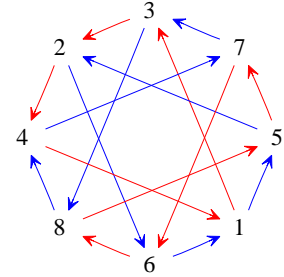}
\caption{
A Cayley graph of $Q_8$.
}
\label{fig.q8}
\end{figure}

Table \ref{tab.Q8c}
shows the index $l$ of the product $g_jg_k=g_l$
at the crossing of row $j$ and column $k$.
In the hypercomplex notation it is summarized as
\begin{equation}
\texttt{-1}^2 = \texttt{1} ;\, \texttt{i}^2 = \texttt{j}^2=\texttt{k}^2=\texttt{-1};
\,
\texttt{i}\cdot \texttt{j} = \texttt{k};
\,
\texttt{j}\cdot \texttt{i} = \texttt{-k};
\label{eq.q8}
\end{equation}
and so on \cite{WildAMM112}.
Fig.\ \ref{fig.q8} is another view, a digraph of 8 vertices, one per
group element \cite{CayleyAJM1}. Blue edges lead from label $j$ to label $l$
if $g_jg_5=g_l$; read edges lead from label $j$ to label $l$
if $g_jg_3=g_l$. (The set of
generators $\{g_5,g_3\}$ here is only one out of many choices.
Equivalent ambivalences govern the Cayley graphs further down.)
\begin{table}
\caption{Cayley (multiplication) table of $Q_{8}$.}
\begin{ruledtabular}
\begin{tabular}{rrrrrrrr}
1& 2& 3& 4& 5& 6& 7& 8\\
2& 1& 4& 3& 6& 5& 8& 7\\
3& 4& 2& 1& 8& 7& 5& 6\\
4& 3& 1& 2& 7& 8& 6& 5\\
5& 6& 7& 8& 2& 1& 4& 3\\
6& 5& 8& 7& 1& 2& 3& 4\\
7& 8& 6& 5& 3& 4& 2& 1\\
8& 7& 5& 6& 4& 3& 1& 2\\
\end{tabular}
\end{ruledtabular}
\label{tab.Q8c}
\end{table}

The 5 conjugacy classes in $Q_8$ are
\begin{eqnarray}
{\mathcal C}_1=\{g_1\},\,
{\mathcal C}_2=\{g_2\},\\
{\mathcal C}_3=\{g_3,g_4\},\,
{\mathcal C}_4=\{g_5,g_6\},\,
{\mathcal C}_5=\{g_7,g_8\}.
\label{eq.C}
\end{eqnarray}

Table \ref{tab.chars} enumerates
the 4 irreducible representations of dimension 1
and 1 of dimension 2 \cite{ScolariciIJTP34}.
\begin{table}
\caption{Characters of the representations $R^{(\alpha)}$ of the five $Q_8$ classes (\ref{eq.C})}
\begin{ruledtabular}
\begin{tabular}{r|rrrrr}
$\backslash \mathcal C$ & 1 & 2 & 3 & 4 & 5 \\
\hline
$R^{(1)}$ & 1 &1 &1 &1 &1 \\
$R^{(2)}$ & 1 &1 &1 &-1 &-1 \\
$R^{(3)}$ & 1 &1 &-1 &1 &-1 \\
$R^{(4)}$ & 1 &1 &-1 &-1 &1 \\
$R^{(5)}$ & 2 &-2 &0 &0 &0 \\
\end{tabular}
\end{ruledtabular}
\label{tab.chars}
\end{table}
The 1-dimensional
representations are already part of the character table,
and the 2-dimensional
representation may be chosen as
\begin{eqnarray}
g_1\equiv \left(\begin{array}{cc}
1 & 0 \\
0 & 1 
\end{array}\right)
;
\quad
g_2\equiv \left(\begin{array}{cc}
-1 & 0 \\
0 & -1
\end{array}\right)
;
\\
g_3 = -g_4\equiv \left(\begin{array}{cc}
i & 0 \\
0 & -i 
\end{array}\right)
;\quad
g_5 = - g_6 \equiv \left(\begin{array}{cc}
0 & -1 \\
1 & 0 
\end{array}\right)
;\quad
\\
g_7= -g_8\equiv \left(\begin{array}{cc}
0 & i \\
i & 0 
\end{array}\right)
\end{eqnarray}
or as the Pauli matrices.

\subsection{CG coefficients}
The Kronecker product of two 1-dimensional representations is obviously
1-dimensional, irreducible and defines CG coefficients (CGS's)
which are all equal to zero or one:
\begin{eqnarray}
R^{(1)}\otimes R^{(\alpha)}=R^{(\alpha)} \quad(\forall \alpha);
\quad
R^{(2)}\otimes R^{(3)} =R^{(4)} ;
\\
R^{2}\otimes R^{(4)} =R^{(3)} ;
\quad
R^{(3)}\otimes R^{(4)} =R^{(2)} .
\end{eqnarray}
The remaining case is the CG series of the square of the 2-dimensional representation,
\begin{equation}
R^{(5)}\otimes R^{(5)}= R^{(1)}\dotplus R^{(2)}\dotplus R^{(3)}\dotplus R^{(4)}.
\end{equation}
(Real-valued 4-dimensional representations like this one are often preferred
in mechanics \cite{GeTASM120}.)
The multiplicities $m_\gamma$
\begin{equation}
R^{(\alpha)}\otimes R^{(\beta)} \equiv \dot \sum_\gamma m_\gamma R^{(\gamma)}
\end{equation}
have been calculated from the traces (characters) $\chi$
of the associated representations via the standard relation
\cite[(3.20)]{Tinkham}
\cite[(7.22)]{ChenRMP57}
\begin{equation}
m_\gamma =\frac{1}{|Q_8|}\sum_g \chi_{\alpha\otimes \beta}(g) \overline{ \chi_\gamma(g)},
\end{equation}
where the overbar denotes complex-conjugation.
(These Kronecker products are called \emph{simply} reducible
because
no multiplicity $m_\gamma$ is larger than one
\cite{WignerAmJM63}.)

The Clebsch--Gordan coefficients $\langle \alpha i,\beta k|\gamma l \rangle$ are the elements of the square
matrix
which provides the similarity transformation of the matrix $R^{(\alpha\otimes \beta)}$ to block diagonal
form \cite{WignerSIAM25}. Its row and column dimension is $n_\alpha n_\beta$, the product
of the dimensions of the matrices $R^{(\alpha)}$ and $R^{(\beta)}$.
An implicit representation is
\cite{KosterPR109,RykhlCPC174}\cite[(7.34b)]{ChenRMP57}
\begin{eqnarray}
\langle \alpha j ,\beta k|\gamma l\rangle
\overline{\langle \alpha j' ,\beta k'|\gamma l'\rangle}
\nonumber
\\
\propto
\frac{1}{|Q_8|}
\sum_g R_{j,j'}^{(\alpha)}(g) R_{k,k'}^{(\beta)}(g) \overline{ R_{l,l'}^{(\gamma)}(g)}
.
\end{eqnarray}
As written here, the summation over $m_\gamma$ replicas of the irreducible representation $R^{(\gamma)}$
is not yet carried out.
In the tables \ref{tab.cg8_5_5}, \ref{tab.cg16_5_5}, \ref{tab.cg16_5_6} and
so on further below, the column double-indices $j$ and $k$ refer to the
representations
in the (essentially arbitrary) ordering
$\alpha \le \beta$
of the representations. The indices $l$ are added to header rows
if the dimensions $n_\gamma$ are larger than one.

We regard this equation as a matrix equation between two matrices with rows
represented by the multi-index $j,k,l$ and columns represented by the multi-index
$j',k',l'$, i.e, matrices with row and column dimension which are the square
of the dimension of the CG matrix. Given the elements of the right hand side
in terms of the elements of the elements of the irreducible representations $R$,
the left hand side is the spectral representation known as Mercer's theorem \cite{BrislawnPAMS104}.
The CG matrices are the eigenvectors of the matrix seen on the right hand side
associated with the non-vanishing eigenvalues \cite{GabrielJMP10_1932}.

All CG matrices tabulated in this manuscript are unitary.

\begin{table}
\caption{CG matrix of $R^{(5)}\otimes R^{(5)}$ of $Q_8$.}
\begin{ruledtabular}
\begin{tabular}{rr|c|c|c|c}
$j$ & $k$ &    $R^{(1)}$ & $R^{(2)}$ & $R^{(3)}$ & $R^{(4)}$ \\
\hline
1 & 1 &    0 & 0 & $1/\surd 2$ & $-1/\surd 2$ \\
1 & 2 &    $-1/\surd 2$ & $1/\surd 2$ & 0 & 0 \\
2 & 1 &    $1/\surd 2$ & $1/\surd 2$ & 0 & 0 \\
2 & 2 &    0 & 0 & $1/\surd 2$ & $1/\surd 2$ \\
\end{tabular}
\end{ruledtabular}
\label{tab.cg8_5_5}
\end{table}
\section{Quaternion Group $Q_{16}$} 
The product table of the $|Q_{16}|=16$ elements
of $Q_{16}$ is
reproduced in Table \ref{tab.Q16c}\@.
\begin{table}
\caption{Cayley table of $Q_{16}$}
\begin{ruledtabular}
\begin{tabular}{rrrrrrrrrrrrrrrr}
 1& 2& 3& 4& 5& 6& 7& 8& 9& 10& 11& 12& 13& 14& 15& 16\\
 2& 1& 4& 3& 6& 5& 8& 7& 10& 9& 12& 11& 14& 13& 16& 15\\
 3& 4& 2& 1& 8& 7& 5& 6& 12& 11& 9& 10& 15& 16& 14& 13\\
 4& 3& 1& 2& 7& 8& 6& 5& 11& 12& 10& 9& 16& 15& 13& 14\\
 5& 6& 7& 8& 2& 1& 4& 3& 15& 16& 14& 13& 11& 12& 10& 9\\
 6& 5& 8& 7& 1& 2& 3& 4& 16& 15& 13& 14& 12& 11& 9& 10\\
 7& 8& 6& 5& 3& 4& 2& 1& 13& 14& 15& 16& 10& 9& 12& 11\\
 8& 7& 5& 6& 4& 3& 1& 2& 14& 13& 16& 15& 9& 10& 11& 12\\
 9& 10& 11& 12& 13& 14& 15& 16& 2& 1& 4& 3& 6& 5& 8& 7\\
 10& 9& 12& 11& 14& 13& 16& 15& 1& 2& 3& 4& 5& 6& 7& 8\\
 11& 12& 10& 9& 16& 15& 13& 14& 3& 4& 2& 1& 8& 7& 5& 6\\
 12& 11& 9& 10& 15& 16& 14& 13& 4& 3& 1& 2& 7& 8& 6& 5\\
 13& 14& 15& 16& 10& 9& 12& 11& 8& 7& 5& 6& 4& 3& 1& 2\\
 14& 13& 16& 15& 9& 10& 11& 12& 7& 8& 6& 5& 3& 4& 2& 1\\
 15& 16& 14& 13& 11& 12& 10& 9& 6& 5& 8& 7& 1& 2& 3& 4\\
 16& 15& 13& 14& 12& 11& 9& 10& 5& 6& 7& 8& 2& 1& 4& 3\\
\end{tabular}
\end{ruledtabular}
\label{tab.Q16c}
\end{table}
The upper left corner is
the $Q_8$ subgroup,
a copy of Table \ref{tab.Q8c}.
(The subgroup structure is discussed by Bohanon and Reid \cite{BohanonJAC23}.)
The Hall--Senior number is 16.14.

\begin{table*}
\caption{Irreducible representations for a subset of $Q_{16}$ members.}
\begin{ruledtabular}
\begin{tabular}{l|cccccccccccccccc}
 & 1 & 2 & 3 & 5 & 7 & 9 & 11 & 13 & 15\\
\hline
$R^{(1)}$ & 1 & 1 & 1 & 1 & 1 & 1 & 1 & 1 & 1 & \\$R^{(2)}$ & 1 & 1 & 1 & 1 & 
1 & -1 & -1 & -1 & -1 & \\$R^{(3)}$ & 1 & 1 & 1 & -1 & -1 & 1 & 1 & -1 & 
-1 & \\$R^{(4)}$ & 1 & 1 & 1 & -1 & -1 & -1 & -1 & 1 & 1 & \\$R^{(
5)}$ & $\left(\begin{array}{cc}1 & 0\\0 & 
1\\\end{array}\right)$ & $\left(\begin{array}{cc}1 & 0\\0 & 
1\\\end{array}\right)$ & $\left(\begin{array}{cc}-1 & 0\\0 & 
-1\\\end{array}\right)$ & $\left(\begin{array}{cc}1 & 0\\0 & 
-1\\\end{array}\right)$ & $\left(\begin{array}{cc}-1 & 0\\0 & 
1\\\end{array}\right)$ & $\left(\begin{array}{cc}0 & 1\\1 & 
0\\\end{array}\right)$ & $\left(\begin{array}{cc}0 & -1\\-1 & 
0\\\end{array}\right)$ & $\left(\begin{array}{cc}0 & -1\\1 & 
0\\\end{array}\right)$ & $\left(\begin{array}{cc}0 & 1\\-1 & 
0\\\end{array}\right)$ & \\$R^{(6)}$ & $\left(\begin{array}{cc}1 & 0\\0 & 
1\\\end{array}\right)$ & $\left(\begin{array}{cc}-1 & 0\\0 & 
-1\\\end{array}\right)$ & $\left(\begin{array}{cc}i & 0\\
0 & -i\\\end{array}\right)$ & $\left(\begin{array}{cc}0 & -1\\1 & 
0\\\end{array}\right)$ & $\left(\begin{array}{cc}0 & i\\i & 
0\\\end{array}\right)$ & $\left(\begin{array}{cc}0 & e_8^3\\e_8 & 
0\\\end{array}\right)$ & $\left(\begin{array}{cc}0 & e_8\\e_8^3 & 
0\\\end{array}\right)$ & $\left(\begin{array}{cc}e_8^3 & 0\\
0 & -e_8\\\end{array}\right)$ & $\left(\begin{array}{cc}-e_8 & 0\\
0 & e_8^3\\\end{array}\right)$ & \\$R^{(7)}$ & $\left(\begin{array}{cc}1 & 0\\
0 & 1\\\end{array}\right)$ & $\left(\begin{array}{cc}-1 & 0\\0 & 
-1\\\end{array}\right)$ & $\left(\begin{array}{cc}i & 0\\
0 & -i\\\end{array}\right)$ & $\left(\begin{array}{cc}0 & -1\\1 & 
0\\\end{array}\right)$ & $\left(\begin{array}{cc}0 & i\\i & 
0\\\end{array}\right)$ & $\left(\begin{array}{cc}0 & -e_8^3\\-e_8 & 
0\\\end{array}\right)$ & $\left(\begin{array}{cc}0 & -e_8\\-e_8^3 & 
0\\\end{array}\right)$ & $\left(\begin{array}{cc}-e_8^3 & 0\\
0 & e_8\\\end{array}\right)$ & $\left(\begin{array}{cc}e_8 & 0\\
0 & -e_8^3\\\end{array}\right)$ &
\end{tabular}
\end{ruledtabular}
\label{tab.Q16irr}
\end{table*}

\begin{figure}
\includegraphics[scale=0.38]{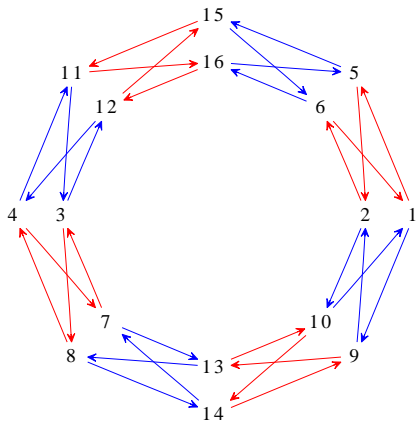}
\caption{
A Cayley graph of $Q_{16}$.
}
\label{fig.q16}
\end{figure}

Fig.\ \ref{fig.q16} is a digraph with vertices labeled by the
indices of the group elements. Blue edges point from
$j$ to $l$ if $g_jg_9=g_l$;
red edges point from
$j$ to $l$ if $g_jg_5=g_l$.

The 7 conjugacy classes are
\begin{eqnarray}
{\mathcal C}_1=\{g_1\},\,
{\mathcal C}_2=\{g_2\},\,
{\mathcal C}_3=\{g_3,g_4\},\,
{\mathcal C}_4=\{g_5,\ldots,g_8\},
\\
{\mathcal C}_5=\{g_9,\ldots,g_{12}\},\,
{\mathcal C}_6=\{g_{13},g_{15}\},\,
{\mathcal C}_7=\{g_{14},g_{16}\},
\end{eqnarray}
with elements of order 1 in ${\mathcal C}_1$,
order 2 in ${\mathcal C}_2$,
order 4 in ${\mathcal C}_3$--${\mathcal C}_5$,
and
order 8 in ${\mathcal C}_6$ and ${\mathcal C}_7$.
There are 7 irreducible representations of dimension 1 or 2 \cite{GAP4,DixonMC24,BabaiMC55}.
Table \ref{tab.Q16irr} shows some of the matrix representations;
the notation $e_k=\exp(2\pi i/k)$ for the $k$-th root
of unity is used.
Representations of the remaining elements $g_k$ of even index $k$ can
be quickly generated
from the neighbor $g_{k-1}$ via
\begin{equation}
g_k=g_2g_{k-1},\quad (k\, \mathrm{even}, Q_8, Q_{16},\, \mathrm{or}\, Q_{32})
\label{eq.gk_1}
\end{equation}
as implied by Table \ref{tab.Q16c}.

Kronecker products involving the 1-dimensional representations
do not split (in the spectroscopic sense). They are already block-diagonal
upon creation
and the CGC can be set to 1:
\begin{eqnarray}
R^{(\alpha)}\otimes R^{(\alpha)}= R^{(1)},\, (\alpha=1,2,3,4);
\\
R^{(2)}\otimes R^{(3)}= R^{(4)};\quad
R^{(2)}\otimes R^{(4)}= R^{(3)};
\\
R^{(\alpha)}\otimes R^{(5)}= R^{(5)},\, (\alpha=1,2,3,4);
\\
R^{(2)}\otimes R^{(6)}= R^{(7)};\quad
R^{(2)}\otimes R^{(7)}= R^{(6)};
\\
R^{(3)}\otimes R^{(4)}= R^{(2)};\quad
R^{(3)}\otimes R^{(6)}= R^{(7)};
\\
R^{(3)}\otimes R^{(7)}= R^{(6)};\quad
R^{(4)}\otimes R^{(6)}= R^{(6)};
\\
R^{(4)}\otimes R^{(7)}= R^{(7)}.
\end{eqnarray}
Kronecker products of 2-dimensional factors
split with multiplicities equal to 1:
\begin{eqnarray}
R^{(5)}\otimes R^{(5)}= R^{(1)}\dotplus R^{(2)}\dotplus R^{(3)}\dotplus R^{(4)};
\label{eq.cg16_5_5}
\\
R^{(5)}\otimes R^{(6)}= R^{(6)}\dotplus R^{(7)};
\\
R^{(5)}\otimes R^{(7)}= R^{(6)}\dotplus R^{(7)};
\\
R^{(6)}\otimes R^{(6)}= R^{(1)}\dotplus R^{(4)}\dotplus R^{(5)};
\\
R^{(6)}\otimes R^{(7)}= R^{(2)}\dotplus R^{(3)}\dotplus R^{(5)};
\\
R^{(7)}\otimes R^{(7)}= R^{(1)}\dotplus R^{(4)}\dotplus R^{(5)}.
\label{eq.cg16_7_7}
\end{eqnarray}
The $4\times 4$ transformation matrices of (\ref{eq.cg16_5_5})--(\ref{eq.cg16_7_7}) are shown
in Tables \ref{tab.cg16_5_5}--\ref{tab.cg16_6_7}.
For non-abelian groups, matrix representations may not be symmetric; 
the freedom of a phase choice remains, but
eigenvectors are not necessarily real-valued.
\begin{table}
\caption{CG matrix of $R^{(5)}\otimes R^{(5)}$ of $Q_{16}$ and of $Q_{32}$.}
\begin{ruledtabular}
\begin{tabular}{rr|cccc}
$j$ & $k$ &    $R^{(1)}$ & $R^{(2)}$ & $R^{(3)}$ & $R^{(4)}$ \\
\hline
1 & 1 &    $1/\surd 2$ & $-1/\surd 2$ & 0 & 0 \\
1 & 2 &    0 & 0 & $1/\surd 2$ & $-1/\surd 2$ \\
2 & 1 &    0 & 0 & $1/\surd 2$ & $1/\surd 2$ \\
2 & 2 &    $1/\surd 2$ & $1/\surd 2$ & 0 & 0 \\
\end{tabular}
\end{ruledtabular}
\label{tab.cg16_5_5}
\end{table}
\begin{table}
\caption{CGC of $R^{(5)}\otimes R^{(6)}$ of $Q_{16}$.
The CGC for 
$R^{(5)}\otimes R^{(7)}$ are the complex-conjugate of these.
}
\begin{ruledtabular}
\begin{tabular}{rr|cc|cc}
    &     &    \multicolumn{2}{c}{$R^{(6)}$} & \multicolumn{2}{c}{$R^{(7)}$} \\
$j$ & $k$ &    1 & 2 & 1 & 2 \\
\hline
1 & 1 &    0 & $i/\surd 2$ & 0 & $-i/\surd 2$\\
1 & 2 &    $-i/\surd 2$ & 0 & $i/\surd 2$ & 0 \\
2 & 1 &    0 & $1/\surd 2$  & 0 & $1/\surd 2$ \\
2 & 2 &    $1/\surd 2$ & 0 & $ 1/\surd 2$ & 0 \\
\end{tabular}
\end{ruledtabular}
\label{tab.cg16_5_6}
\end{table}
\begin{table}
\caption{CGC of $R^{(6)}\otimes R^{(6)}$
or
$R^{(7)}\otimes R^{(7)}$
of $Q_{16}$.
}
\begin{ruledtabular}
\begin{tabular}{rr|c|c|cc}
    &     &    $R^{(1)}$ & $R^{(4)}$ & \multicolumn{2}{c}{$R^{(5)}$} \\
$j$ & $k$ &      &   & 1 & 2 \\
\hline
1 & 1 &    0 & 0 & $-i\surd 2$ & $-1/\surd 2$ \\
1 & 2 &    $-1/\surd 2$ & $1/\surd 2$ & 0 & 0\\
2 & 1 &    $1/\surd 2$ & $1/\surd 2$ & 0 & 0\\
2 & 2 &    0 & 0 & $-i\surd 2$ & $1/\surd 2$ \\
\end{tabular}
\end{ruledtabular}
\label{tab.cg16_6_6}
\end{table}
\begin{table}
\caption{CGC of $R^{(6)}\otimes R^{(7)}$ of $Q_{16}$.
}
\begin{ruledtabular}
\begin{tabular}{rr|c|c|cc}
    &     &    $R^{(2)}$ & $R^{(3)}$ & \multicolumn{2}{c}{$R^{(5)}$} \\
$j$ & $k$ &      &   & 1 & 2 \\
\hline
1 & 1 &    0 & 0 & $i\surd 2$ & $-1/\surd 2$ \\
1 & 2 &    $-1/\surd 2$ & $1/\surd 2$ & 0 & 0\\
2 & 1 &    $1/\surd 2$ & $1/\surd 2$ & 0 & 0\\
2 & 2 &    0 & 0 & $i\surd 2$ & $1/\surd 2$ \\
\end{tabular}
\end{ruledtabular}
\label{tab.cg16_6_7}
\end{table}

\section{Quaternion Group $Q_{32}$}
The multiplication table of the quaternion group of order
$|Q_{32}|=32$ is Table \ref{tab.Q32c}.
\begin{table*}[tb]
\caption{Cayley table of $Q_{32}$, Hall--Senior group 32.51}
\begin{ruledtabular}
\begin{tabular}{rrrrrrrrrrrrrrrrrrrrrrrrrrrrrrrr}
1& 2& 3& 4& 5& 6& 7& 8& 9& 10& 11& 12& 13& 14& 15& 16& 17& 18& 19& 20& 21& 
22& 23& 24& 25& 26& 27& 28& 29& 30& 31& 32\\
2& 1& 4& 3& 6& 5& 8& 7& 10& 9& 12& 11& 14& 13& 16& 15& 18& 17& 20& 19& 22& 
21& 24& 23& 26& 25& 28& 27& 30& 29& 32& 31\\
3& 4& 2& 1& 7& 8& 6& 5& 12& 11& 9& 10& 16& 15& 13& 14& 20& 19& 17& 18& 24& 
23& 21& 22& 27& 28& 26& 25& 31& 32& 30& 29\\
4& 3& 1& 2& 8& 7& 5& 6& 11& 12& 10& 9& 15& 16& 14& 13& 19& 20& 18& 17& 23& 
24& 22& 21& 28& 27& 25& 26& 32& 31& 29& 30\\
5& 6& 7& 8& 4& 3& 1& 2& 15& 16& 14& 13& 9& 10& 11& 12& 23& 24& 22& 21& 17& 
18& 19& 20& 29& 30& 31& 32& 28& 27& 25& 26\\
6& 5& 8& 7& 3& 4& 2& 1& 16& 15& 13& 14& 10& 9& 12& 11& 24& 23& 21& 22& 18& 
17& 20& 19& 30& 29& 32& 31& 27& 28& 26& 25\\
7& 8& 6& 5& 1& 2& 3& 4& 13& 14& 15& 16& 12& 11& 9& 10& 21& 22& 23& 24& 20& 
19& 17& 18& 31& 32& 30& 29& 25& 26& 27& 28\\
8& 7& 5& 6& 2& 1& 4& 3& 14& 13& 16& 15& 11& 12& 10& 9& 22& 21& 24& 23& 19& 
20& 18& 17& 32& 31& 29& 30& 26& 25& 28& 27\\
9& 10& 11& 12& 13& 14& 15& 16& 2& 1& 4& 3& 6& 5& 8& 7& 29& 30& 31& 32& 28& 
27& 25& 26& 24& 23& 21& 22& 18& 17& 20& 19\\
10& 9& 12& 11& 14& 13& 16& 15& 1& 2& 3& 4& 5& 6& 7& 8& 30& 29& 32& 31& 27& 
28& 26& 25& 23& 24& 22& 21& 17& 18& 19& 20\\
11& 12& 10& 9& 15& 16& 14& 13& 3& 4& 2& 1& 7& 8& 6& 5& 32& 31& 29& 30& 26& 
25& 28& 27& 21& 22& 23& 24& 20& 19& 17& 18\\
12& 11& 9& 10& 16& 15& 13& 14& 4& 3& 1& 2& 8& 7& 5& 6& 31& 32& 30& 29& 25& 
26& 27& 28& 22& 21& 24& 23& 19& 20& 18& 17\\
13& 14& 15& 16& 12& 11& 9& 10& 8& 7& 5& 6& 2& 1& 4& 3& 25& 26& 27& 28& 29& 
30& 31& 32& 18& 17& 20& 19& 22& 21& 24& 23\\
14& 13& 16& 15& 11& 12& 10& 9& 7& 8& 6& 5& 1& 2& 3& 4& 26& 25& 28& 27& 30& 
29& 32& 31& 17& 18& 19& 20& 21& 22& 23& 24\\
15& 16& 14& 13& 9& 10& 11& 12& 6& 5& 8& 7& 3& 4& 2& 1& 28& 27& 25& 26& 32& 
31& 29& 30& 20& 19& 17& 18& 24& 23& 21& 22\\
16& 15& 13& 14& 10& 9& 12& 11& 5& 6& 7& 8& 4& 3& 1& 2& 27& 28& 26& 25& 31& 
32& 30& 29& 19& 20& 18& 17& 23& 24& 22& 21\\
17& 18& 19& 20& 21& 22& 23& 24& 25& 26& 27& 28& 29& 30& 31& 32& 2& 1& 4& 3& 
6& 5& 8& 7& 10& 9& 12& 11& 14& 13& 16& 15\\
18& 17& 20& 19& 22& 21& 24& 23& 26& 25& 28& 27& 30& 29& 32& 31& 1& 2& 3& 4& 
5& 6& 7& 8& 9& 10& 11& 12& 13& 14& 15& 16\\
19& 20& 18& 17& 23& 24& 22& 21& 28& 27& 25& 26& 32& 31& 29& 30& 3& 4& 2& 1& 
7& 8& 6& 5& 12& 11& 9& 10& 16& 15& 13& 14\\
20& 19& 17& 18& 24& 23& 21& 22& 27& 28& 26& 25& 31& 32& 30& 29& 4& 3& 1& 2& 
8& 7& 5& 6& 11& 12& 10& 9& 15& 16& 14& 13\\
21& 22& 23& 24& 20& 19& 17& 18& 31& 32& 30& 29& 25& 26& 27& 28& 8& 7& 5& 6& 
2& 1& 4& 3& 14& 13& 16& 15& 11& 12& 10& 9\\
22& 21& 24& 23& 19& 20& 18& 17& 32& 31& 29& 30& 26& 25& 28& 27& 7& 8& 6& 5& 
1& 2& 3& 4& 13& 14& 15& 16& 12& 11& 9& 10\\
23& 24& 22& 21& 17& 18& 19& 20& 29& 30& 31& 32& 28& 27& 25& 26& 6& 5& 8& 7& 
3& 4& 2& 1& 16& 15& 13& 14& 10& 9& 12& 11\\
24& 23& 21& 22& 18& 17& 20& 19& 30& 29& 32& 31& 27& 28& 26& 25& 5& 6& 7& 8& 
4& 3& 1& 2& 15& 16& 14& 13& 9& 10& 11& 12\\
25& 26& 27& 28& 29& 30& 31& 32& 18& 17& 20& 19& 22& 21& 24& 23& 14& 13& 16& 
15& 11& 12& 10& 9& 7& 8& 6& 5& 1& 2& 3& 4\\
26& 25& 28& 27& 30& 29& 32& 31& 17& 18& 19& 20& 21& 22& 23& 24& 13& 14& 15& 
16& 12& 11& 9& 10& 8& 7& 5& 6& 2& 1& 4& 3\\
27& 28& 26& 25& 31& 32& 30& 29& 19& 20& 18& 17& 23& 24& 22& 21& 15& 16& 14& 
13& 9& 10& 11& 12& 6& 5& 8& 7& 3& 4& 2& 1\\
28& 27& 25& 26& 32& 31& 29& 30& 20& 19& 17& 18& 24& 23& 21& 22& 16& 15& 13& 
14& 10& 9& 12& 11& 5& 6& 7& 8& 4& 3& 1& 2\\
29& 30& 31& 32& 28& 27& 25& 26& 24& 23& 21& 22& 18& 17& 20& 19& 10& 9& 12& 
11& 14& 13& 16& 15& 1& 2& 3& 4& 5& 6& 7& 8\\
30& 29& 32& 31& 27& 28& 26& 25& 23& 24& 22& 21& 17& 18& 19& 20& 9& 10& 11& 
12& 13& 14& 15& 16& 2& 1& 4& 3& 6& 5& 8& 7\\
31& 32& 30& 29& 25& 26& 27& 28& 22& 21& 24& 23& 19& 20& 18& 17& 11& 12& 10& 
9& 15& 16& 14& 13& 3& 4& 2& 1& 7& 8& 6& 5\\
32& 31& 29& 30& 26& 25& 28& 27& 21& 22& 23& 24& 20& 19& 17& 18& 12& 11& 9& 
10& 16& 15& 13& 14& 4& 3& 1& 2& 8& 7& 5& 6\\
\end{tabular}
\end{ruledtabular}
\label{tab.Q32c}
\end{table*}

\begin{figure}[hbt]
\includegraphics[scale=0.5]{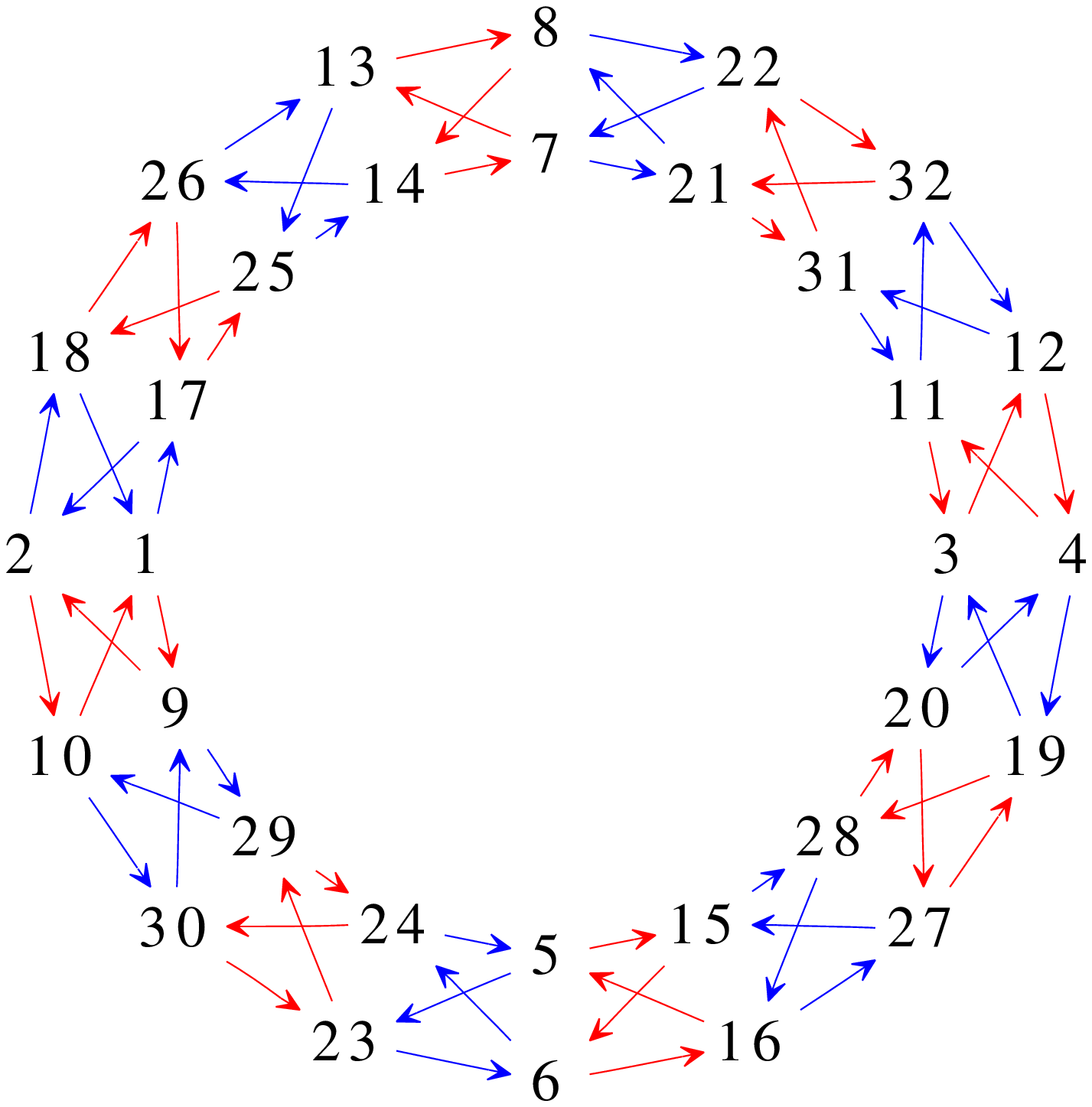}
\caption{
A Cayley graph of $Q_{32}$.
Blue edges represent multiplication by $g_{17}$ and
red edges represent multiplication by $g_9$.
}
\label{fig.q32}
\end{figure}

The group comprises 11 conjugacy classes:
\begin{eqnarray}
{\mathcal C}_1= \{g_1\},\,
{\mathcal C}_2=\{g_2\},\,
{\mathcal C}_3=\{g_3,g_4\},\,
{\mathcal C}_4=\{g_5,g_7\},
\\
{\mathcal C}_5= \{g_6,g_8\},\,
{\mathcal C}_6=\{g_9,g_{10},\ldots,g_{16}\},
\\
{\mathcal C}_7= \{g_{17},g_{18},\ldots,g_{24}\},\,
{\mathcal C}_8=\{g_{25},g_{29}\},
\\
{\mathcal C}_9= \{g_{26},g_{30}\},\,
{\mathcal C}_{10}=\{g_{27},g_{32}\},\,
{\mathcal C}_{11}=\{g_{28},g_{31}\}.
\end{eqnarray}
The 4 one-dimensional and 7 two-dimensional irreducible representations
are partially shown in Table \ref{tab.Q32irr}. \emph{Partially} means
that the representations for the remaining members of the group can be easily
calculated with the aid of Table \ref{tab.Q32c}.
A glance at Figure \ref{fig.q32} reveals that following the blue edges
and the red edges one can indeed generate
all elements starting at $g_1$; all but the columns labeled $9$ and $17$
in Table \ref{tab.Q32irr} are redundant to that task.

\begin{turnpage}
\begin{table}
\caption{Irreducible representations for some $Q_{32}$ elements.}
\begin{ruledtabular}
\begin{tabular}{l|cccccccccccccccc}
 & 2 & 3 & 5 & 6 & 9 & 17 & 25 & 26 & 27 & 28\\ \hline
$R^{(1)}$ & 1 & 1 & 1 & 1 & 1 & 1 & 1 & 1 & 1 & 1 & \\$R^{(2)}$ & 1 & 1 & 1 & 
1 & 1 & -1 & -1 & -1 & -1 & -1 & \\$R^{(3)}$ & 1 & 1 & 1 & 1 & -1 & 1 & -1 & 
-1 & -1 & -1 & \\$R^{(4)}$ & 1 & 1 & 1 & 1 & -1 & -1 & 1 & 1 & 1 & 1 & \\$R^{(
5)}$ & $\left(\begin{array}{cc}1 & 0\\0 & 
1\\\end{array}\right)$ & $\left(\begin{array}{cc}1 & 0\\0 & 
1\\\end{array}\right)$ & $\left(\begin{array}{cc}-1 & 0\\0 & 
-1\\\end{array}\right)$ & $\left(\begin{array}{cc}-1 & 0\\0 & 
-1\\\end{array}\right)$ & $\left(\begin{array}{cc}1 & 0\\0 & 
-1\\\end{array}\right)$ & $\left(\begin{array}{cc}0 & 1\\1 & 
0\\\end{array}\right)$ & $\left(\begin{array}{cc}0 & -1\\1 & 
0\\\end{array}\right)$ & $\left(\begin{array}{cc}0 & -1\\1 & 
0\\\end{array}\right)$ & $\left(\begin{array}{cc}0 & -1\\1 & 
0\\\end{array}\right)$ & $\left(\begin{array}{cc}0 & -1\\1 & 
0\\\end{array}\right)$ & \\$R^{(6)}$ & $\left(\begin{array}{cc}1 & 0\\0 & 
1\\\end{array}\right)$ & $\left(\begin{array}{cc}-1 & 0\\0 & 
-1\\\end{array}\right)$ & $\left(\begin{array}{cc}i & 0\\
0 & -i\\\end{array}\right)$ & $\left(\begin{array}{cc}i & 0\\
0 & -i\\\end{array}\right)$ & $\left(\begin{array}{cc}0 & 1\\1 & 
0\\\end{array}\right)$ & $\left(\begin{array}{cc}0 & e_8^3\\-e_8 & 
0\\\end{array}\right)$ & $\left(\begin{array}{cc}e_8^3 & 0\\
0 & -e_8\\\end{array}\right)$ & $\left(\begin{array}{cc}e_8^3 & 0\\
0 & -e_8\\\end{array}\right)$ & $\left(\begin{array}{cc}-e_8^3 & 0\\
0 & e_8\\\end{array}\right)$ & $\left(\begin{array}{cc}-e_8^3 & 0\\
0 & e_8\\\end{array}\right)$ & \\$R^{(7)}$ & $\left(\begin{array}{cc}1 & 0\\
0 & 1\\\end{array}\right)$ & $\left(\begin{array}{cc}-1 & 0\\0 & 
-1\\\end{array}\right)$ & $\left(\begin{array}{cc}i & 0\\
0 & -i\\\end{array}\right)$ & $\left(\begin{array}{cc}i & 0\\
0 & -i\\\end{array}\right)$ & $\left(\begin{array}{cc}0 & 1\\1 & 
0\\\end{array}\right)$ & $\left(\begin{array}{cc}0 & -e_8^3\\e_8 & 
0\\\end{array}\right)$ & $\left(\begin{array}{cc}-e_8^3 & 0\\
0 & e_8\\\end{array}\right)$ & $\left(\begin{array}{cc}-e_8^3 & 0\\
0 & e_8\\\end{array}\right)$ & $\left(\begin{array}{cc}e_8^3 & 0\\
0 & -e_8\\\end{array}\right)$ & $\left(\begin{array}{cc}e_8^3 & 0\\
0 & -e_8\\\end{array}\right)$ & \\$R^{(8)}$ & $\left(\begin{array}{cc}-1 & 0\\
0 & -1\\\end{array}\right)$ & $\left(\begin{array}{cc}i & 0\\
0 & -i\\\end{array}\right)$ & $\left(\begin{array}{cc}e_8^3 & 0\\
0 & -e_8\\\end{array}\right)$ & $\left(\begin{array}{cc}-e_8^3 & 0\\
0 & e_8\\\end{array}\right)$ & $\left(\begin{array}{cc}0 & -1\\1 & 
0\\\end{array}\right)$ & $\left(\begin{array}{cc}0 & e_{16}^5\\e_{16}^3 & 
0\\\end{array}\right)$ & $\left(\begin{array}{cc}e_{16}^5 & 0\\0 & 
-e_{16}^3\\\end{array}\right)$ & $\left(\begin{array}{cc}-e_{16}^5 & 0\\0 & 
e_{16}^3\\\end{array}\right)$ & $\left(\begin{array}{cc}-e_{16} & 0\\0 & 
e_{16}^7\\\end{array}\right)$ & $\left(\begin{array}{cc}e_{16} & 0\\0 & 
-e_{16}^7\\\end{array}\right)$ & \\$R^{(9)}$ & $\left(\begin{array}{cc}-1 & 0\\
0 & -1\\\end{array}\right)$ & $\left(\begin{array}{cc}i & 0\\
0 & -i\\\end{array}\right)$ & $\left(\begin{array}{cc}e_8^3 & 0\\
0 & -e_8\\\end{array}\right)$ & $\left(\begin{array}{cc}-e_8^3 & 0\\
0 & e_8\\\end{array}\right)$ & $\left(\begin{array}{cc}0 & -1\\1 & 
0\\\end{array}\right)$ & $\left(\begin{array}{cc}0 & -e_{16}^5\\-e_{16}^3 & 
0\\\end{array}\right)$ & $\left(\begin{array}{cc}-e_{16}^5 & 0\\0 & 
e_{16}^3\\\end{array}\right)$ & $\left(\begin{array}{cc}e_{16}^5 & 0\\0 & 
-e_{16}^3\\\end{array}\right)$ & $\left(\begin{array}{cc}e_{16} & 0\\0 & 
-e_{16}^7\\\end{array}\right)$ & $\left(\begin{array}{cc}-e_{16} & 0\\0 & 
e_{16}^7\\\end{array}\right)$ & \\$R^{(10)}$ & $\left(\begin{array}{cc}-1 & 0\\
0 & -1\\\end{array}\right)$ & $\left(\begin{array}{cc}i & 0\\
0 & -i\\\end{array}\right)$ & $\left(\begin{array}{cc}-e_8^3 & 0\\
0 & e_8\\\end{array}\right)$ & $\left(\begin{array}{cc}e_8^3 & 0\\
0 & -e_8\\\end{array}\right)$ & $\left(\begin{array}{cc}0 & -1\\1 & 
0\\\end{array}\right)$ & $\left(\begin{array}{cc}0 & -e_{16}\\-e_{16}^7 & 
0\\\end{array}\right)$ & $\left(\begin{array}{cc}-e_{16} & 0\\0 & 
e_{16}^7\\\end{array}\right)$ & $\left(\begin{array}{cc}e_{16} & 0\\0 & 
-e_{16}^7\\\end{array}\right)$ & $\left(\begin{array}{cc}-e_{16}^5 & 0\\0 & 
e_{16}^3\\\end{array}\right)$ & $\left(\begin{array}{cc}e_{16}^5 & 0\\0 & 
-e_{16}^3\\\end{array}\right)$ & \\$R^{(11)}$ & $\left(\begin{array}{cc}-1 & 
0\\0 & -1\\\end{array}\right)$ & $\left(\begin{array}{cc}i & 0\\
0 & -i\\\end{array}\right)$ & $\left(\begin{array}{cc}-e_8^3 & 0\\
0 & e_8\\\end{array}\right)$ & $\left(\begin{array}{cc}e_8^3 & 0\\
0 & -e_8\\\end{array}\right)$ & $\left(\begin{array}{cc}0 & -1\\1 & 
0\\\end{array}\right)$ & $\left(\begin{array}{cc}0 & e_{16}\\e_{16}^7 & 
0\\\end{array}\right)$ & $\left(\begin{array}{cc}e_{16} & 0\\0 & 
-e_{16}^7\\\end{array}\right)$ & $\left(\begin{array}{cc}-e_{16} & 0\\0 & 
e_{16}^7\\\end{array}\right)$ & $\left(\begin{array}{cc}e_{16}^5 & 0\\0 & 
-e_{16}^3\\\end{array}\right)$ & $\left(\begin{array}{cc}-e_{16}^5 & 0\\0 & 
e_{16}^3\\\end{array}\right)$ & \\
\end{tabular}
\end{ruledtabular}
\label{tab.Q32irr}
\end{table}
\end{turnpage}
\clearpage

The non-splitting CG series are:
\begin{eqnarray}
R^{(\alpha)}\otimes R^{(\alpha)}= R^{(1)};\,(\alpha=2,3,4);
\\
R^{(2)}\otimes R^{(3)}= R^{(4)};
\\
R^{(2)}\otimes R^{(4)}= R^{(3)};
\\
R^{(\alpha)}\otimes R^{(5)}= R^{(5)};\,(\alpha=2,3,4);
\\
R^{(\alpha)}\otimes R^{(6)}= R^{(7)};\, 
R^{(\alpha)}\otimes R^{(7)}= R^{(6)};\,(\alpha=2,3);
\\
R^{(2)}\otimes R^{(8)}= R^{(9)};\,
R^{(2)}\otimes R^{(9)}= R^{(8)};
\\
R^{(2)}\otimes R^{(10)}= R^{(11)};\,
R^{(2)}\otimes R^{(11)}= R^{(10)};
\end{eqnarray}
\begin{eqnarray}
R^{(3)}\otimes R^{(4)}= R^{(2)};\quad
R^{(3)}\otimes R^{(8)}= R^{(9)};
\\
R^{(3)}\otimes R^{(9)}= R^{(8)};\quad
R^{(3)}\otimes R^{(10)}= R^{(11)};
\\
R^{(3)}\otimes R^{(11)}= R^{(10)};
\\
R^{(4)}\otimes R^{(k)}= R^{(k)};\,(k=6,7,\ldots,11)
\end{eqnarray}
The splitting series are:
\begin{eqnarray}
R^{(5)}\otimes R^{(5)}= R^{(1)}\dotplus R^{(2)}\dotplus R^{(3)}\dotplus R^{(4)};
\label{eq.q32_5_5}
\\
R^{(5)}\otimes R^{(6)}= R^{(6)}\dotplus R^{(7)};
\label{eq.q32_5_6}
\\
R^{(5)}\otimes R^{(7)}= R^{(6)}\dotplus R^{(7)};
\\
R^{(5)}\otimes R^{(8)}= R^{(10)}\dotplus R^{(11)};
\\
R^{(5)}\otimes R^{(9)}= R^{(10)}\dotplus R^{(11)};
\\
R^{(5)}\otimes R^{(10)}= R^{(8)}\dotplus R^{(9)};
\\
R^{(5)}\otimes R^{(11)}= R^{(8)}\dotplus R^{(9)};
\end{eqnarray}
\begin{eqnarray}
R^{(6)}\otimes R^{(6)}= R^{(1)}\dotplus R^{(4)}\dotplus R^{(5)};
\\
R^{(6)}\otimes R^{(7)}= R^{(2)}\dotplus R^{(3)}\dotplus R^{(5)};
\\
R^{(6)}\otimes R^{(8)}= R^{(8)}\dotplus R^{(11)};
\\
R^{(6)}\otimes R^{(9)}= R^{(9)}\dotplus R^{(10)};
\\
R^{(6)}\otimes R^{(10)}= R^{(9)}\dotplus R^{(11)};
\\
R^{(6)}\otimes R^{(11)}= R^{(8)}\dotplus R^{(10)};
\end{eqnarray}
\begin{eqnarray}
R^{(7)}\otimes R^{(7)}= R^{(1)}\dotplus R^{(4)}\dotplus R^{(5)};
\\
R^{(7)}\otimes R^{(8)}= R^{(9)}\dotplus R^{(10)};
\\
R^{(7)}\otimes R^{(9)}= R^{(8)}\dotplus R^{(11)};
\\
R^{(7)}\otimes R^{(10)}= R^{(8)}\dotplus R^{(10)};
\\
R^{(7)}\otimes R^{(11)}= R^{(9)}\dotplus R^{(11)};
\end{eqnarray}
\begin{eqnarray}
R^{(8)}\otimes R^{(8)}= R^{(1)}\dotplus R^{(4)}\dotplus R^{(6)};
\\
R^{(8)}\otimes R^{(9)}= R^{(2)}\dotplus R^{(3)}\dotplus R^{(7)};
\\
R^{(8)}\otimes R^{(10)}= R^{(5)}\dotplus R^{(7)};
\\
R^{(8)}\otimes R^{(11)}= R^{(5)}\dotplus R^{(6)};
\end{eqnarray}
\begin{eqnarray}
R^{(9)}\otimes R^{(9)}= R^{(1)}\dotplus R^{(4)}\dotplus R^{(6)};
\\
R^{(9)}\otimes R^{(10)}= R^{(5)}\dotplus R^{(6)};
\\
R^{(9)}\otimes R^{(11)}= R^{(5)}\dotplus R^{(7)};
\end{eqnarray}
\begin{eqnarray}
R^{(10)}\otimes R^{(10)}= R^{(1)}\dotplus R^{(4)}\dotplus R^{(7)};
\\
R^{(10)}\otimes R^{(11)}= R^{(2)}\dotplus R^{(3)}\dotplus R^{(6)};
\\
R^{(11)}\otimes R^{(11)}= R^{(1)}\dotplus R^{(4)}\dotplus R^{(7)}.
\label{eq.q32_11_11}
\end{eqnarray}

The case (\ref{eq.q32_5_5}) is covered
by Table \ref{tab.cg16_5_5}, the cases
(\ref{eq.q32_5_6})--(\ref{eq.q32_11_11}) by Tables \ref{tab.cg32_5_6}--\ref{tab.cg32_8_11}.

\begin{table}
\caption{CGC of $R^{(5)}\otimes R^{(6)}$ of $Q_{32}$.
The coefficients of
$R^{(5)}\otimes R^{(7)}$ are the complex-conjugate of these.
}
\begin{ruledtabular}
\begin{tabular}{rr|cc|cc}
    &     &    \multicolumn{2}{c}{$R^{(6)}$} & \multicolumn{2}{c}{$R^{(7)}$} \\
$j$ & $k$ &    1 & 2 & 1 & 2 \\
\hline
1 & 1 &    0 & $i/\surd 2$ & 0 & $-i/\surd 2$\\
1 & 2 &    $i/\surd 2$ & 0 & $-i/\surd 2$ & 0 \\
2 & 1 &    0 & $1/\surd 2$  & 0 & $1/\surd 2$ \\
2 & 2 &    $-1/\surd 2$ & 0 & $-1/\surd 2$ & 0 \\
\end{tabular}
\end{ruledtabular}
\label{tab.cg32_5_6}
\end{table}
\begin{table}
\caption{CGC of $R^{(5)}\otimes R^{(8)}$ of $Q_{32}$.
The coefficients of
$R^{(5)}\otimes R^{(9)}$ are the complex-conjugate of these.
}
\begin{ruledtabular}
\begin{tabular}{rr|cc|cc}
    &     &    \multicolumn{2}{c}{$R^{(10)}$} & \multicolumn{2}{c}{$R^{(11)}$} \\
$j$ & $k$ &    1 & 2 & 1 & 2 \\
\hline
1 & 1 &    $-i/\surd 2$ & 0 & $i/\surd 2$ & 0\\
1 & 2 &    0 &  $-i/\surd 2$ & 0 & $i/\surd 2$ \\
2 & 1 &    $-1/\surd 2$  & 0 & $-1/\surd 2$ & 0\\
2 & 2 &    0 & $1/\surd 2$ & 0 & $1/\surd 2$ \\
\end{tabular}
\end{ruledtabular}
\label{tab.cg32_5_8}
\end{table}
\begin{table}
\caption{CGC of $R^{(5)}\otimes R^{(10)}$ of $Q_{32}$.
The coefficients of
$R^{(5)}\otimes R^{(11)}$ are the complex-conjugate of these.
}
\begin{ruledtabular}
\begin{tabular}{rr|cc|cc}
    &     &    \multicolumn{2}{c}{$R^{(8)}$} & \multicolumn{2}{c}{$R^{(9)}$} \\
$j$ & $k$ &    1 & 2 & 1 & 2 \\
\hline
1 & 1 &    $i/\surd 2$ & 0 & $-i/\surd 2$ & 0\\
1 & 2 &    0 &  $i/\surd 2$ & 0 & $-i/\surd 2$ \\
2 & 1 &    $-1/\surd 2$  & 0 & $-1/\surd 2$ & 0\\
2 & 2 &    0 & $1/\surd 2$ & 0 & $1/\surd 2$ \\
\end{tabular}
\end{ruledtabular}
\label{tab.cg32_5_10}
\end{table}
\begin{table}
\caption{CG matrix of $R^{(6)}\otimes R^{(6)}$
or $R^{(7)}\otimes R^{(7)}$
of $Q_{32}$.
}
\begin{ruledtabular}
\begin{tabular}{rr|cc|cc}
    &     &    $R^{(1)}$ & $R^{(4)}$ & \multicolumn{2}{c}{$R^{(5)}$} \\
$j$ & $k$ &     &  & 1 & 2 \\
\hline
1 & 1 &    0 & 0 & $-i/\surd 2$ & $-1/\surd 2$ \\
1 & 2 &    $1/\surd 2$ & $-1/\surd 2$ & 0 & 0\\
2 & 1 &    $1/\surd 2$ & $1/\surd 2$ & 0 & 0\\
2 & 2 &    0 & 0 & $-i/\surd 2$ & $1/\surd 2$ \\
\end{tabular}
\end{ruledtabular}
\label{tab.cg32_6_6}
\end{table}
\begin{table}
\caption{CGC of $R^{(6)}\otimes R^{(7)}$ of $Q_{32}$.
}
\begin{ruledtabular}
\begin{tabular}{rr|c|c|cc}
    &     &    $R^{(2)}$ & $R^{(3)}$ & \multicolumn{2}{c}{$R^{(5)}$} \\
$j$ & $k$ &      &   & 1 & 2 \\
\hline
1 & 1 &    0 & 0 & $i\surd 2$ & $-1/\surd 2$ \\
1 & 2 &    $1/\surd 2$ & $-1/\surd 2$ & 0 & 0\\
2 & 1 &    $1/\surd 2$ & $1/\surd 2$ & 0 & 0\\
2 & 2 &    0 & 0 & $i\surd 2$ & $1/\surd 2$ \\
\end{tabular}
\end{ruledtabular}
\label{tab.cg32_6_7}
\end{table}

\begin{table}
\caption{CGC of $R^{(6)}\otimes R^{(8)}$
or $R^{(7)}\otimes R^{(9)}$
of $Q_{32}$.
}
\begin{ruledtabular}
\begin{tabular}{rr|cc|cc}
    &     &    \multicolumn{2}{c}{$R^{(8)}$} & \multicolumn{2}{c}{$R^{(11)}$} \\
$j$ & $k$ &    1 & 2 & 1 & 2 \\
\hline
1 & 1 &    0 & 1 & 0 & 0\\
1 & 2 &    0 &  0 & -1 & 0 \\
2 & 1 &    0 & 0 & 0 & 1 \\
2 & 2 &    -1 & 0 & 0 & 0 \\
\end{tabular}
\end{ruledtabular}
\label{tab.cg32_6_8}
\end{table}

\begin{table}
\caption{CGC of $R^{(6)}\otimes R^{(9)}$
or $R^{(7)}\otimes R^{(8)}$
of $Q_{32}$.
}
\begin{ruledtabular}
\begin{tabular}{rr|cc|cc}
    &     &    \multicolumn{2}{c}{$R^{(9)}$} & \multicolumn{2}{c}{$R^{(10)}$} \\
$j$ & $k$ &    1 & 2 & 1 & 2 \\
\hline
1 & 1 &    0 & 1 & 0 & 0\\
1 & 2 &    0 &  0 & -1 & 0 \\
2 & 1 &    0 & 0 & 0 & 1 \\
2 & 2 &    -1 & 0 & 0 & 0 \\
\end{tabular}
\end{ruledtabular}
\label{tab.cg32_6_9}
\end{table}

\begin{table}
\caption{CGC of $R^{(6)}\otimes R^{(10)}$
or $R^{(7)}\otimes R^{(11)}$
of $Q_{32}$.
}
\begin{ruledtabular}
\begin{tabular}{rr|cc|cc}
    &     &    \multicolumn{2}{c}{$R^{(9)}$} & \multicolumn{2}{c}{$R^{(11)}$} \\
$j$ & $k$ &    1 & 2 & 1 & 2 \\
\hline
1 & 1 &    0 & 0 & 0 & 1\\
1 & 2 &    -1 &  0 & 0 & 0 \\
2 & 1 &    0 & 1 & 0 & 0 \\
2 & 2 &    0 & 0 & -1 & 0 \\
\end{tabular}
\end{ruledtabular}
\label{tab.cg32_6_10}
\end{table}

\begin{table}
\caption{CGC of $R^{(6)}\otimes R^{(11)}$
or
$R^{(7)}\otimes R^{(10)}$
of $Q_{32}$.
}
\begin{ruledtabular}
\begin{tabular}{rr|cc|cc}
    &     &    \multicolumn{2}{c}{$R^{(8)}$} & \multicolumn{2}{c}{$R^{(10)}$} \\
$j$ & $k$ &    1 & 2 & 1 & 2 \\
\hline
1 & 1 &    0 & 0 & 0 & 1\\
1 & 2 &    -1 &  0 & 0 & 0 \\
2 & 1 &    0 & 1 & 0 & 0 \\
2 & 2 &    0 & 0 & -1 & 0 \\
\end{tabular}
\end{ruledtabular}
\label{tab.cg32_6_11}
\end{table}

\begin{table}
\caption{CG matrix of $R^{(8)}\otimes R^{(8)}$
or $R^{(9)}\otimes R^{(9)}$
of $Q_{32}$.
Replacing $R^{(6)}$ by $R^{(7)}$ provides the table
for $R^{(10)}\otimes R^{(10)}$ and $R^{(11)}\otimes R^{(10)}$.
}
\begin{ruledtabular}
\begin{tabular}{rr|cc|cc}
    &     &    $R^{(1)}$ & $R^{(4)}$ & \multicolumn{2}{c}{$R^{(6)}$} \\
$j$ & $k$ &             &           & 1 & 2 \\
\hline
1 & 1 &    0 & 0 & 0 & 1 \\
1 & 2 &    $-1/\surd 2$ & $1/\surd 2$ & 0 & 0\\
2 & 1 &    $1/\surd 2$ & $1/\surd 2$ & 0 & 0\\
2 & 2 &    0 & 0 & 1 & 0 \\
\end{tabular}
\end{ruledtabular}
\label{tab.cg32_8_8}
\end{table}

\begin{table}
\caption{CG matrix of $R^{(8)}\otimes R^{(9)}$ of $Q_{32}$.
Replacing $R^{(7)}$ by $R^{(6)}$ provides the table
for $R^{(10)}\otimes R^{(11)}$.
}
\begin{ruledtabular}
\begin{tabular}{rr|cc|cc}
    &     &    $R^{(2)}$ & $R^{(3)}$ & \multicolumn{2}{c}{$R^{(7)}$} \\
$j$ & $k$ &     &  &  1 & 2 \\
\hline
1 & 1 &    0 & 0 & 0 & 1 \\
1 & 2 &    $-1/\surd 2$ & $1/\surd 2$ & 0 & 0\\
2 & 1 &    $1/\surd 2$ & $1/\surd 2$ & 0 & 0\\
2 & 2 &    0 & 0 & 1 & 0 \\
\end{tabular}
\end{ruledtabular}
\label{tab.cg32_8_9}
\end{table}

\begin{table}
\caption{CGC of $R^{(8)}\otimes R^{(10)}$
or $R^{(9)}\otimes R^{(11)}$ 
of $Q_{32}$.
}
\begin{ruledtabular}
\begin{tabular}{rr|cc|cc}
    &     &    \multicolumn{2}{c}{$R^{(5)}$} & \multicolumn{2}{c}{$R^{(7)}$} \\
$j$ & $k$ &    1 & 2 & 1 & 2 \\
\hline
1 & 1 &    0 & 0 & 1 & 0\\
1 & 2 &    $i/\surd 2$ &  $1/\surd 2$ & 0 & 0 \\
2 & 1 &    $-i/\surd 2$ & $1/\surd 2$ & 0 & 0 \\
2 & 2 &    0 & 0 & 0 & 1 \\
\end{tabular}
\end{ruledtabular}
\label{tab.cg32_8_10}
\end{table}

\begin{table}
\caption{CGC of $R^{(8)}\otimes R^{(11)}$
or $R^{(9)}\otimes R^{(10)}$
of $Q_{32}$.
}
\begin{ruledtabular}
\begin{tabular}{rr|cc|cc}
    &     &    \multicolumn{2}{c}{$R^{(5)}$} & \multicolumn{2}{c}{$R^{(6)}$} \\
$j$ & $k$ &    1 & 2 & 1 & 2 \\
\hline
1 & 1 &    0 & 0 & 1 & 0\\
1 & 2 &    $-i/\surd 2$ &  $1/\surd 2$ & 0 & 0 \\
2 & 1 &    $i/\surd 2$ & $1/\surd 2$ & 0 & 0 \\
2 & 2 &    0 & 0 & 0 & 1 \\
\end{tabular}
\end{ruledtabular}
\label{tab.cg32_8_11}
\end{table}

\section{($C_2\times D_8)\rtimes C_2$}
The factor group of $Q_8\times Q_8$ with respect to the subgroup
containing the unit and $\texttt{-1}\times \texttt{-1}$ builds another group
of order 32, the 49th out of the 51 groups of order 32 in the
Small Group Library \cite{GAP4}, known as $\Gamma_5 a_1$ \cite{Hall}
and named 32.42 hereafter \cite{HalesJA316,Thomas,ZimmermGMJ41}.
It can be assembled by a direct and semi-direct product
of the cyclic group $C_2$
and the dihedral group $D_8$
\cite{floretion,ThyssenJCP129}.
The elements shall be sorted according
to the multiplication table \ref{tab.floretc}.
\begin{table*}
\caption{Cayley table of $32.42$}
\begin{ruledtabular}
\begin{tabular}{rrrrrrrrrrrrrrrrrrrrrrrrrrrrrrrr}
1& 2& 3& 4& 5& 6& 7& 8& 9& 10& 11& 12& 13& 14& 15& 16& 17& 18& 19& 20& 21& 
22& 23& 24& 25& 26& 27& 28& 29& 30& 31& 32\\
2& 1& 4& 3& 6& 5& 8& 7& 10& 9& 12& 11& 14& 13& 16& 15& 18& 17& 20& 19& 22& 
21& 24& 23& 26& 25& 28& 27& 30& 29& 32& 31\\
3& 4& 1& 2& 7& 8& 5& 6& 11& 12& 9& 10& 15& 16& 13& 14& 20& 19& 18& 17& 24& 
23& 22& 21& 28& 27& 26& 25& 32& 31& 30& 29\\
4& 3& 2& 1& 8& 7& 6& 5& 12& 11& 10& 9& 16& 15& 14& 13& 19& 20& 17& 18& 23& 
24& 21& 22& 27& 28& 25& 26& 31& 32& 29& 30\\
5& 6& 7& 8& 1& 2& 3& 4& 14& 13& 16& 15& 10& 9& 12& 11& 21& 22& 23& 24& 17& 
18& 19& 20& 30& 29& 32& 31& 26& 25& 28& 27\\
6& 5& 8& 7& 2& 1& 4& 3& 13& 14& 15& 16& 9& 10& 11& 12& 22& 21& 24& 23& 18& 
17& 20& 19& 29& 30& 31& 32& 25& 26& 27& 28\\
7& 8& 5& 6& 3& 4& 1& 2& 16& 15& 14& 13& 12& 11& 10& 9& 24& 23& 22& 21& 20& 
19& 18& 17& 31& 32& 29& 30& 27& 28& 25& 26\\
8& 7& 6& 5& 4& 3& 2& 1& 15& 16& 13& 14& 11& 12& 9& 10& 23& 24& 21& 22& 19& 
20& 17& 18& 32& 31& 30& 29& 28& 27& 26& 25\\
9& 10& 11& 12& 13& 14& 15& 16& 1& 2& 3& 4& 5& 6& 7& 8& 26& 25& 28& 27& 30& 
29& 32& 31& 18& 17& 20& 19& 22& 21& 24& 23\\
10& 9& 12& 11& 14& 13& 16& 15& 2& 1& 4& 3& 6& 5& 8& 7& 25& 26& 27& 28& 29& 
30& 31& 32& 17& 18& 19& 20& 21& 22& 23& 24\\
11& 12& 9& 10& 15& 16& 13& 14& 3& 4& 1& 2& 7& 8& 5& 6& 27& 28& 25& 26& 31& 
32& 29& 30& 19& 20& 17& 18& 23& 24& 21& 22\\
12& 11& 10& 9& 16& 15& 14& 13& 4& 3& 2& 1& 8& 7& 6& 5& 28& 27& 26& 25& 32& 
31& 30& 29& 20& 19& 18& 17& 24& 23& 22& 21\\
13& 14& 15& 16& 9& 10& 11& 12& 6& 5& 8& 7& 2& 1& 4& 3& 30& 29& 32& 31& 26& 
25& 28& 27& 21& 22& 23& 24& 17& 18& 19& 20\\
14& 13& 16& 15& 10& 9& 12& 11& 5& 6& 7& 8& 1& 2& 3& 4& 29& 30& 31& 32& 25& 
26& 27& 28& 22& 21& 24& 23& 18& 17& 20& 19\\
15& 16& 13& 14& 11& 12& 9& 10& 8& 7& 6& 5& 4& 3& 2& 1& 31& 32& 29& 30& 27& 
28& 25& 26& 24& 23& 22& 21& 20& 19& 18& 17\\
16& 15& 14& 13& 12& 11& 10& 9& 7& 8& 5& 6& 3& 4& 1& 2& 32& 31& 30& 29& 28& 
27& 26& 25& 23& 24& 21& 22& 19& 20& 17& 18\\
17& 18& 19& 20& 21& 22& 23& 24& 25& 26& 27& 28& 29& 30& 31& 32& 1& 2& 3& 4& 
5& 6& 7& 8& 9& 10& 11& 12& 13& 14& 15& 16\\
18& 17& 20& 19& 22& 21& 24& 23& 26& 25& 28& 27& 30& 29& 32& 31& 2& 1& 4& 3& 
6& 5& 8& 7& 10& 9& 12& 11& 14& 13& 16& 15\\
19& 20& 17& 18& 23& 24& 21& 22& 27& 28& 25& 26& 31& 32& 29& 30& 4& 3& 2& 1& 
8& 7& 6& 5& 12& 11& 10& 9& 16& 15& 14& 13\\
20& 19& 18& 17& 24& 23& 22& 21& 28& 27& 26& 25& 32& 31& 30& 29& 3& 4& 1& 2& 
7& 8& 5& 6& 11& 12& 9& 10& 15& 16& 13& 14\\
21& 22& 23& 24& 17& 18& 19& 20& 30& 29& 32& 31& 26& 25& 28& 27& 5& 6& 7& 8& 
1& 2& 3& 4& 14& 13& 16& 15& 10& 9& 12& 11\\
22& 21& 24& 23& 18& 17& 20& 19& 29& 30& 31& 32& 25& 26& 27& 28& 6& 5& 8& 7& 
2& 1& 4& 3& 13& 14& 15& 16& 9& 10& 11& 12\\
23& 24& 21& 22& 19& 20& 17& 18& 32& 31& 30& 29& 28& 27& 26& 25& 8& 7& 6& 5& 
4& 3& 2& 1& 15& 16& 13& 14& 11& 12& 9& 10\\
24& 23& 22& 21& 20& 19& 18& 17& 31& 32& 29& 30& 27& 28& 25& 26& 7& 8& 5& 6& 
3& 4& 1& 2& 16& 15& 14& 13& 12& 11& 10& 9\\
25& 26& 27& 28& 29& 30& 31& 32& 17& 18& 19& 20& 21& 22& 23& 24& 10& 9& 12& 
11& 14& 13& 16& 15& 2& 1& 4& 3& 6& 5& 8& 7\\
26& 25& 28& 27& 30& 29& 32& 31& 18& 17& 20& 19& 22& 21& 24& 23& 9& 10& 11& 
12& 13& 14& 15& 16& 1& 2& 3& 4& 5& 6& 7& 8\\
27& 28& 25& 26& 31& 32& 29& 30& 19& 20& 17& 18& 23& 24& 21& 22& 11& 12& 9& 
10& 15& 16& 13& 14& 3& 4& 1& 2& 7& 8& 5& 6\\
28& 27& 26& 25& 32& 31& 30& 29& 20& 19& 18& 17& 24& 23& 22& 21& 12& 11& 10& 
9& 16& 15& 14& 13& 4& 3& 2& 1& 8& 7& 6& 5\\
29& 30& 31& 32& 25& 26& 27& 28& 22& 21& 24& 23& 18& 17& 20& 19& 14& 13& 16& 
15& 10& 9& 12& 11& 5& 6& 7& 8& 1& 2& 3& 4\\
30& 29& 32& 31& 26& 25& 28& 27& 21& 22& 23& 24& 17& 18& 19& 20& 13& 14& 15& 
16& 9& 10& 11& 12& 6& 5& 8& 7& 2& 1& 4& 3\\
31& 32& 29& 30& 27& 28& 25& 26& 24& 23& 22& 21& 20& 19& 18& 17& 15& 16& 13& 
14& 11& 12& 9& 10& 8& 7& 6& 5& 4& 3& 2& 1\\
32& 31& 30& 29& 28& 27& 26& 25& 23& 24& 21& 22& 19& 20& 17& 18& 16& 15& 14& 
13& 12& 11& 10& 9& 7& 8& 5& 6& 3& 4& 1& 2\\
\end{tabular}
\end{ruledtabular}
\label{tab.floretc}
\end{table*}

\begin{figure}
\includegraphics[scale=0.55]{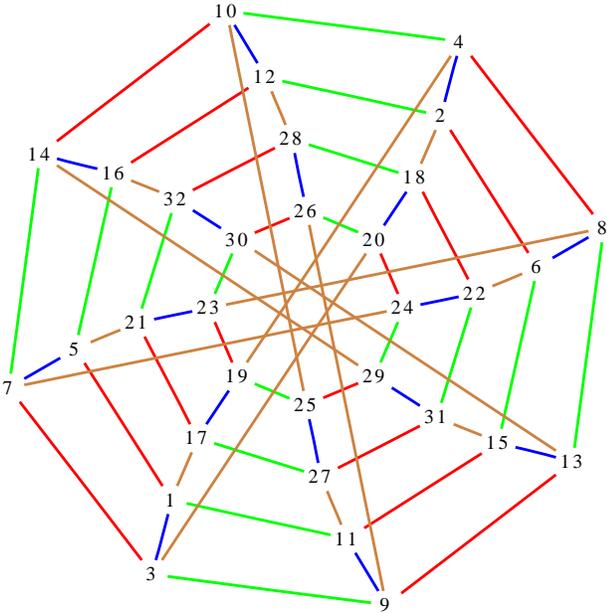}
\caption{
A Cayley graph of 32.42.
}
\label{fig.q3242}
\end{figure}

If the multiplications with the generators $g_3$, $g_5$, $g_{11}$
and $g_{17}$ are coded with blue, red, green and brown edges, respectively,
the Cayley graph in Fig.\ \ref{fig.q3242} results.
(These four generators are self-inverse, so the edges
are undirected.)

\clearpage

With this enumeration of the elements, the 17 conjugacy classes
are
\begin{eqnarray}
{\mathcal C}_1=\{g_1\},
{\mathcal C}_2=\{g_2\},\\
{\mathcal C}_j=\{g_{2j-3},g_{2j-2}\},
(j=3,4,\ldots, 17).
\label{eq.cflorCj}
\end{eqnarray}
The 16 one-dimensional representations are part of Table \ref{tab.florirr}.
\begin{table}
\caption{Characters of the 32.42 classes (\ref{eq.cflorCj}).}
\begin{ruledtabular}
\begin{tabular}{l|ccccccccccccccccc}
$\backslash \cal C$& 1 & 2 & 3 & 4 & 5 & 6 & 7 & 8 & 9 & 10 & 11 & 12 & 13 & 14 & 15 & 16 & 17\\ \hline
$R^{(1)}$ & 1 & 1 & 1 & 1 & 1 & 1 & 1 & 1 & 1 & 1 & 1 & 1 & 1 & 1 & 1 & 1 & 1
\\
$R^{(2)}$ & 1 & 1 & 1 & 1 & 1 & 1 & 1 & 1 & 1 & -1 & -1 & -1 & -1 & -1 & -1 & 
-1 & -1
\\
$R^{(3)}$ & 1 & 1 & 1 & 1 & 1 & -1 & -1 & -1 & -1 & 1 & 1 & 1 & 1 & -1 & -1 & 
-1 & -1
\\
$R^{(4)}$ & 1 & 1 & 1 & 1 & 1 & -1 & -1 & -1 & -1 & -1 & -1 & -1 & -1 & 1 & 
1 & 1 & 1
\\
$R^{(5)}$ & 1 & 1 & 1 & -1 & -1 & 1 & 1 & -1 & -1 & 1 & 1 & -1 & -1 & 1 & 1 & 
-1 & -1
\\
$R^{(6)}$ & 1 & 1 & 1 & -1 & -1 & 1 & 1 & -1 & -1 & -1 & -1 & 1 & 1 & -1 & 
-1 & 1 & 1
\\
$R^{(7)}$ & 1 & 1 & 1 & -1 & -1 & -1 & -1 & 1 & 1 & 1 & 1 & -1 & -1 & -1 & 
-1 & 1 & 1
\\
$R^{(8)}$ & 1 & 1 & 1 & -1 & -1 & -1 & -1 & 1 & 1 & -1 & -1 & 1 & 1 & 1 & 1 & 
-1 & -1
\\
$R^{(9)}$ & 1 & 1 & -1 & 1 & -1 & 1 & -1 & 1 & -1 & 1 & -1 & 1 & -1 & 1 & 
-1 & 1 & -1
\\
$R^{(10)}$ & 1 & 1 & -1 & 1 & -1 & 1 & -1 & 1 & -1 & -1 & 1 & -1 & 1 & -1 & 
1 & -1 & 1
\\
$R^{(11)}$ & 1 & 1 & -1 & 1 & -1 & -1 & 1 & -1 & 1 & 1 & -1 & 1 & -1 & -1 & 
1 & -1 & 1
\\
$R^{(12)}$ & 1 & 1 & -1 & 1 & -1 & -1 & 1 & -1 & 1 & -1 & 1 & -1 & 1 & 1 & 
-1 & 1 & -1
\\
$R^{(13)}$ & 1 & 1 & -1 & -1 & 1 & 1 & -1 & -1 & 1 & 1 & -1 & -1 & 1 & 1 & 
-1 & -1 & 1
\\
$R^{(14)}$ & 1 & 1 & -1 & -1 & 1 & 1 & -1 & -1 & 1 & -1 & 1 & 1 & -1 & -1 & 
1 & 1 & -1
\\
$R^{(15)}$ & 1 & 1 & -1 & -1 & 1 & -1 & 1 & 1 & -1 & 1 & -1 & -1 & 1 & -1 & 
1 & 1 & -1
\\
$R^{(16)}$ & 1 & 1 & -1 & -1 & 1 & -1 & 1 & 1 & -1 & -1 & 1 & 1 & -1 & 1 & 
-1 & -1 & 1
\\
$R^{(17)}$ & 4 & -4 & 0 & 0 & 0 & 0 & 0 & 0 & 0 & 0 & 0 & 0 & 0 & 0 & 0 & 0 & 
0
\\
\end{tabular}
\end{ruledtabular}
\label{tab.florirr}
\end{table}
The 4-dimensional representation is the unit and negative
unit matrix for $g_1$ and $g_2$. A possible set of the other elements is
written with the aid of $2\times 2$ sub-matrices
\begin{eqnarray}
\tau_0=\left(\begin{array}{cc} 1 & 0 \\ 0 & 1 \\ \end{array}\right) ;
\tau_1=\left(\begin{array}{cc} 0 & 1 \\ 1 & 0 \\ \end{array}\right) ;
\\
\tau_2=\left(\begin{array}{cc} 0 & -1 \\ 1 & 0 \\ \end{array}\right) ;
\tau_3=\left(\begin{array}{cc} 1 & 0 \\ 0 & -1 \\ \end{array}\right)
\end{eqnarray}
as
\begin{eqnarray}
g_3=\left(\begin{array}{cc} \tau_0 & 0\\0 & -\tau_0 \end{array}\right);
\,
g_5= \left(\begin{array}{cc}\tau_3 & 0 \\0 & \tau_3 \end{array}\right)
\\
g_7=\left(\begin{array}{cc}\tau_3 & 0 \\0 & -\tau_3 \end{array}\right) ;
\,
g_9=\left(\begin{array}{cc}\tau_1 & 0\\0& \tau_1 \end{array}\right);
\end{eqnarray}
\begin{eqnarray}
g_{11}=\left(\begin{array}{cccc}\tau_1 & 0 \\ 0 & -\tau_1 \end{array}\right);
\,
g_{13}=\left(\begin{array}{cccc}\tau_2 & 0 \\ 0 & \tau_2\end{array}\right);
\\
g_{15}=\left(\begin{array}{cc}\tau_2 & 0 \\0 & -\tau_2 \end{array}\right);
\,
g_{17}=\left(\begin{array}{cc}0 & \tau_3 \\ \tau_3 & 0 \end{array}\right);
\end{eqnarray}
\begin{eqnarray}
g_{19}=\left(\begin{array}{cc}0 & -\tau_3 \\ \tau_3 & 0\end{array}\right);
\,
g_{21}=\left(\begin{array}{cc}0 & \tau_0 \\ \tau_0 & 0 \end{array}\right);
\\
g_{23}=\left(\begin{array}{cc}0 & -\tau_0 \\ \tau_0 & 0 \end{array}\right);
\,
g_{25}=\left(\begin{array}{cc}0 & -\tau_2 \\-\tau_2 & 0 \end{array}\right);
\end{eqnarray}
\begin{eqnarray}
g_{27}=\left(\begin{array}{cc}0 & \tau_2 \\ -\tau_2 & 0 \end{array}\right);
\,
g_{29}=\left(\begin{array}{cc}0 & -\tau_1 \\ -\tau_1 & 0 \end{array}\right);
\\
g_{31}=\left(\begin{array}{cc}0 & \tau_1 \\ -\tau_1 & 0 \end{array}\right).
\end{eqnarray}
All remaining representations are available by multiplication
with $g_2$ as indicated by Table \ref{tab.floretc} and Eq.\ (\ref{eq.gk_1}).

The Kronecker products of pairs of one-dimensional representations
are not discussed in detail, because this reduces
to multiplying two rows containing a string of $\pm 1$ in the character table
\ref{tab.florirr} and finding the row that matches that binary pattern.
The cases
\begin{equation}
R^{(\alpha)}\otimes R^{(17)} = R^{(17)}
,\,(\alpha=1,\ldots,16)
\end{equation}
do not split either. The remaining
case is
\begin{equation}
R^{(17)}\otimes R^{(17)}=
R^{(1)}\dotplus
R^{(2)}\dotplus
R^{(3)}\dotplus \cdots
\dotplus
R^{(16)}.
\end{equation}
Its associated CG matrix defines Table \ref{tab.q32_17_17}.
\begin{table*}
\caption{CGC of $R^{(17)}\otimes R^{(17)}$ of 32.42.}
\begin{ruledtabular}
\begin{tabular}{rr|cccccccccccccccc}
$j$    & $k$    & 
$R^{(1)}$ &
$R^{(2)}$ &
$R^{(3)}$ &
$R^{(4)}$ &
$R^{(5)}$ &
$R^{(6)}$ &
$R^{(7)}$ &
$R^{(8)}$ &
$R^{(9)}$ &
$R^{(10)}$ &
$R^{(11)}$ &
$R^{(12)}$ &
$R^{(13)}$ &
$R^{(14)}$ &
$R^{(15)}$ &
$R^{(16)}$ \\
\hline
1 & 1 &    $1/2$ & $-1/2$ & $-1/2$ & $1/2$ & 0 & 0 &0 &0 &0 &0 &0 &0 &0 &0 &0 &0\\
1 & 2 &    0 & 0 & 0 & 0 & $-1/2$ & $1/2$ & $1/2$ & $-1/2$ & 0 & 0 &0 &0 &0 &0 &0 &0 \\
1 & 3 &    0 & 0 & 0 & 0 & 0 & 0& 0& 0& $1/2$ & $-1/2$ & $-1/2$ & $1/2$ & 0 & 0 &0 &0 \\
1 & 4 &    0 & 0 & 0 & 0 & 0 & 0& 0& 0& 0& 0&0&0& $-1/2$ & $1/2$ & $1/2$ & $-1/2$ \\
2 & 1 &    0 & 0 & 0 & 0 & $-1/2$ & $1/2$ & $-1/2$ & $1/2$ & 0 & 0 &0 &0 &0 &0 &0 &0 \\
2 & 2 &    $1/2$ & $-1/2$ & $1/2$ & $-1/2$ & 0 & 0 &0 &0 &0 &0 &0 &0 &0 &0 &0 &0\\
2 & 3 &    0 & 0 & 0 & 0 & 0 & 0& 0& 0& 0& 0&0&0& $-1/2$ & $1/2$ & $-1/2$ & $1/2$ \\
2 & 4 &    0 & 0 & 0 & 0 & 0 & 0& 0& 0& $1/2$ & $-1/2$ & $1/2$ & $-1/2$ & 0 & 0 &0 &0 \\
3 & 1 &    0 & 0 & 0 & 0 & 0 & 0& 0& 0& $1/2$ & $1/2$ & $-1/2$ & $-1/2$ & 0 & 0 &0 &0 \\
3 & 2 &    0 & 0 & 0 & 0 & 0 & 0& 0& 0& 0& 0&0&0& $1/2$ & $1/2$ & $-1/2$ & $-1/2$ \\
3 & 3 &     $1/2$ & $1/2$ & $-1/2$ & $-1/2$ & 0 & 0 &0 &0 &0 &0 &0 &0 &0 &0 &0 &0\\
3 & 4 &    0 & 0 & 0 & 0 & $1/2$ & $1/2$ & $-1/2$ & $-1/2$ & 0 & 0 &0 &0 &0 &0 &0 &0 \\
4 & 1 &    0 & 0 & 0 & 0 & 0 & 0& 0& 0& 0& 0&0&0& $1/2$ & $1/2$ & $1/2$ & $1/2$ \\
4 & 2 &    0 & 0 & 0 & 0 & 0 & 0& 0& 0& $1/2$ & $1/2$ & $1/2$ & $1/2$ & 0 & 0 &0 &0 \\
4 & 3 &    0 & 0 & 0 & 0 & $1/2$ & $1/2$ & $1/2$ & $1/2$ & 0 & 0 &0 &0 &0 &0 &0 &0 \\
4 & 4 &    $1/2$ & $1/2$ & $1/2$ & $1/2$ & 0 & 0 &0 &0 &0 &0 &0 &0 &0 &0 &0 &0\\
\end{tabular}
\end{ruledtabular}
\label{tab.q32_17_17}
\end{table*}

\section{Summary}
The Clebsch--Gordan coefficients of four finite non-abelian groups with
irreducible representations of dimension up to $n_\alpha=4$ have been
computed by direct diagonalization of matrices of dimension $n_\alpha^4$.

\bibliography{all}

\begin{thebibliography}{34}
\expandafter\ifx\csname natexlab\endcsname\relax\def\natexlab#1{#1}\fi
\expandafter\ifx\csname bibnamefont\endcsname\relax
  \def\bibnamefont#1{#1}\fi
\expandafter\ifx\csname bibfnamefont\endcsname\relax
  \def\bibfnamefont#1{#1}\fi
\expandafter\ifx\csname citenamefont\endcsname\relax
  \def\citenamefont#1{#1}\fi
\expandafter\ifx\csname url\endcsname\relax
  \def\url#1{\texttt{#1}}\fi
\expandafter\ifx\csname urlprefix\endcsname\relax\def\urlprefix{URL }\fi
\providecommand{\bibinfo}[2]{#2}
\providecommand{\eprint}[2][]{\url{#2}}

\bibitem[{\citenamefont{Miller}(1898)}]{MillerPAPS37}
\bibinfo{author}{\bibfnamefont{G.~A.} \bibnamefont{Miller}},
  \bibinfo{journal}{Proc. Am. Phil. Soc.} \textbf{\bibinfo{volume}{37}},
  \bibinfo{pages}{312} (\bibinfo{year}{1898}).

\bibitem[{\citenamefont{Coxeter}(1973)}]{CoxeterCPAM26}
\bibinfo{author}{\bibfnamefont{H.~S.~M.} \bibnamefont{Coxeter}},
  \bibinfo{journal}{Comm. Pure Appl. Math.} \textbf{\bibinfo{volume}{26}},
  \bibinfo{pages}{693} (\bibinfo{year}{1973}).

\bibitem[{\citenamefont{Girard}(1984)}]{GirardEJP5}
\bibinfo{author}{\bibfnamefont{P.~R.} \bibnamefont{Girard}},
  \bibinfo{journal}{European J. Phys.} \textbf{\bibinfo{volume}{5}},
  \bibinfo{pages}{25} (\bibinfo{year}{1984}).

\bibitem[{\citenamefont{Saue and Jensen}(1999)}]{SaueJCP111}
\bibinfo{author}{\bibfnamefont{T.}~\bibnamefont{Saue}} \bibnamefont{and}
  \bibinfo{author}{\bibfnamefont{H.~J.~A.} \bibnamefont{Jensen}},
  \bibinfo{journal}{J. Chem. Phys.} \textbf{\bibinfo{volume}{111}},
  \bibinfo{pages}{6211} (\bibinfo{year}{1999}).

\bibitem[{\citenamefont{Visscher}(2002)}]{VisscherJCC23}
\bibinfo{author}{\bibfnamefont{L.}~\bibnamefont{Visscher}},
  \bibinfo{journal}{J. Comput.\ Chem.} \textbf{\bibinfo{volume}{23}},
  \bibinfo{pages}{759} (\bibinfo{year}{2002}).

\bibitem[{\citenamefont{Flurry}(1980)}]{Flurry}
\bibinfo{author}{\bibfnamefont{R.~L.} \bibnamefont{Flurry}},
  \emph{\bibinfo{title}{Symmetry Groups}} (\bibinfo{publisher}{Prentice Hall},
  \bibinfo{address}{Englewood Cliffs, NJ}, \bibinfo{year}{1980}).

\bibitem[{\citenamefont{Koster et~al.}(1963)\citenamefont{Koster, Dimmok,
  Wheeler, and Statz}}]{Koster}
\bibinfo{author}{\bibfnamefont{G.~F.} \bibnamefont{Koster}},
  \bibinfo{author}{\bibfnamefont{J.~O.} \bibnamefont{Dimmok}},
  \bibinfo{author}{\bibfnamefont{R.~G.} \bibnamefont{Wheeler}},
  \bibnamefont{and} \bibinfo{author}{\bibfnamefont{H.}~\bibnamefont{Statz}},
  \emph{\bibinfo{title}{Properties of the thirty-two point groups}}
  (\bibinfo{publisher}{MIT Press}, \bibinfo{address}{Cambridge},
  \bibinfo{year}{1963}).

\bibitem[{\citenamefont{Stringham}(1881)}]{StringhamAJM4}
\bibinfo{author}{\bibfnamefont{W.~I.} \bibnamefont{Stringham}},
  \bibinfo{journal}{Amer. J. Math.} \textbf{\bibinfo{volume}{4}},
  \bibinfo{pages}{345} (\bibinfo{year}{1881}).

\bibitem[{\citenamefont{Besche et~al.}(2002)\citenamefont{Besche, Eick, and
  O'Brien}}]{BescheIJAC12}
\bibinfo{author}{\bibfnamefont{H.~U.} \bibnamefont{Besche}},
  \bibinfo{author}{\bibfnamefont{B.}~\bibnamefont{Eick}}, \bibnamefont{and}
  \bibinfo{author}{\bibfnamefont{E.~A.} \bibnamefont{O'Brien}},
  \bibinfo{journal}{Internat. J. Algebra Comput.}
  \textbf{\bibinfo{volume}{12}}, \bibinfo{pages}{623} (\bibinfo{year}{2002}).

\bibitem[{\citenamefont{Besche and Eick}(1999)}]{BescheJSC27}
\bibinfo{author}{\bibfnamefont{H.~U.} \bibnamefont{Besche}} \bibnamefont{and}
  \bibinfo{author}{\bibfnamefont{B.}~\bibnamefont{Eick}}, \bibinfo{journal}{J.
  Symbolic Comput.} \textbf{\bibinfo{volume}{27}}, \bibinfo{pages}{405}
  (\bibinfo{year}{1999}).

\bibitem[{GAP(2008)}]{GAP4}
\emph{\bibinfo{title}{{GAP} -- Groups, Algorithms, and Programming, Version
  4.4.12}}, \bibinfo{organization}{The {GAP} Group} (\bibinfo{year}{2008}),
  \urlprefix\url{http://www.gap-system.org}.

\bibitem[{\citenamefont{Fieldsteel et~al.}(2008)\citenamefont{Fieldsteel,
  Lindberg, London, Tran, and Xu}}]{FieldsteelASP2008}
\bibinfo{author}{\bibfnamefont{N.~M.} \bibnamefont{Fieldsteel}},
  \bibinfo{author}{\bibfnamefont{T.~J.} \bibnamefont{Lindberg}},
  \bibinfo{author}{\bibfnamefont{T.~A.} \bibnamefont{London}},
  \bibinfo{author}{\bibfnamefont{H.}~\bibnamefont{Tran}}, \bibnamefont{and}
  \bibinfo{author}{\bibfnamefont{H.}~\bibnamefont{Xu}}, in
  \emph{\bibinfo{booktitle}{2008 Arizona Summer Program on Computational Group
  Theory}} (\bibinfo{year}{2008}),
  \urlprefix\url{http://math.arizona.edu/~asp/2008/StrongSymmetricGenusPaper.p%
df}.

\bibitem[{\citenamefont{Schaps}(2010)}]{Schaps}
\bibinfo{author}{\bibfnamefont{M.}~\bibnamefont{Schaps}},
  \emph{\bibinfo{title}{Database of group character tables of solvable groups}}
  (\bibinfo{year}{2010}),
  \urlprefix\url{http://u.cs.biu.ac.il/~mschaps/DATA/database.html}.

\bibitem[{\citenamefont{Hall and Senior}(1964)}]{Hall}
\bibinfo{author}{\bibfnamefont{M.}~\bibnamefont{Hall}} \bibnamefont{and}
  \bibinfo{author}{\bibfnamefont{J.~K.} \bibnamefont{Senior}},
  \emph{\bibinfo{title}{The groups of order $2^n$, $n\le 6$}}
  (\bibinfo{publisher}{MacMillan}, \bibinfo{year}{1964}).

\bibitem[{\citenamefont{Thomas and Wood}(1980)}]{Thomas}
\bibinfo{author}{\bibfnamefont{A.~D.} \bibnamefont{Thomas}} \bibnamefont{and}
  \bibinfo{author}{\bibfnamefont{G.~V.} \bibnamefont{Wood}},
  \emph{\bibinfo{title}{Group Tables}} (\bibinfo{publisher}{Shiva Publishing},
  \bibinfo{year}{1980}).

\bibitem[{\citenamefont{Hales et~al.}(2007)\citenamefont{Hales, Passi, and
  Wilson}}]{HalesJA316}
\bibinfo{author}{\bibfnamefont{A.~W.} \bibnamefont{Hales}},
  \bibinfo{author}{\bibfnamefont{I.~B.~S.} \bibnamefont{Passi}},
  \bibnamefont{and} \bibinfo{author}{\bibfnamefont{L.~E.}
  \bibnamefont{Wilson}}, \bibinfo{journal}{J. Algebra}
  \textbf{\bibinfo{volume}{316}}, \bibinfo{pages}{109} (\bibinfo{year}{2007}).

\bibitem[{\citenamefont{Wild}(2005)}]{WildAMM112}
\bibinfo{author}{\bibfnamefont{M.}~\bibnamefont{Wild}}, \bibinfo{journal}{Amer.
  Math. Monthly} \textbf{\bibinfo{volume}{112}}, \bibinfo{pages}{20}
  (\bibinfo{year}{2005}).

\bibitem[{\citenamefont{Cayley}(1878)}]{CayleyAJM1}
\bibinfo{author}{\bibfnamefont{A.}~\bibnamefont{Cayley}},
  \bibinfo{journal}{Amer. J. Math.} \textbf{\bibinfo{volume}{1}},
  \bibinfo{pages}{174} (\bibinfo{year}{1878}).

\bibitem[{\citenamefont{Scolarici and Solombrino}(1995)}]{ScolariciIJTP34}
\bibinfo{author}{\bibfnamefont{G.}~\bibnamefont{Scolarici}} \bibnamefont{and}
  \bibinfo{author}{\bibfnamefont{L.}~\bibnamefont{Solombrino}},
  \bibinfo{journal}{Int. J. Theor. Phys.} \textbf{\bibinfo{volume}{34}},
  \bibinfo{pages}{2491} (\bibinfo{year}{1995}).

\bibitem[{\citenamefont{Ge}(1998)}]{GeTASM120}
\bibinfo{author}{\bibfnamefont{Q.~J.} \bibnamefont{Ge}},
  \bibinfo{journal}{Trans. AMSE J. Mech. Design}
  \textbf{\bibinfo{volume}{120}}, \bibinfo{pages}{404} (\bibinfo{year}{1998}).

\bibitem[{\citenamefont{Tinkham}(1992)}]{Tinkham}
\bibinfo{author}{\bibfnamefont{M.}~\bibnamefont{Tinkham}},
  \emph{\bibinfo{title}{Group theory and quantum mechanics}}
  (\bibinfo{publisher}{Dover Publications}, \bibinfo{year}{1992}),
  \bibinfo{edition}{9th} ed., ISBN \bibinfo{isbn}{0-486-43247-5}.

\bibitem[{\citenamefont{Chen et~al.}(1985)\citenamefont{Chen, Gao, and
  Ma}}]{ChenRMP57}
\bibinfo{author}{\bibfnamefont{J.-Q.} \bibnamefont{Chen}},
  \bibinfo{author}{\bibfnamefont{M.-J.} \bibnamefont{Gao}}, \bibnamefont{and}
  \bibinfo{author}{\bibfnamefont{G.-Q.} \bibnamefont{Ma}},
  \bibinfo{journal}{Rev. Mod. Phys.} \textbf{\bibinfo{volume}{57}},
  \bibinfo{pages}{211} (\bibinfo{year}{1985}).

\bibitem[{\citenamefont{Wigner}(1941)}]{WignerAmJM63}
\bibinfo{author}{\bibfnamefont{E.~P.} \bibnamefont{Wigner}},
  \bibinfo{journal}{Amer. J. Math.} \textbf{\bibinfo{volume}{63}},
  \bibinfo{pages}{57} (\bibinfo{year}{1941}).

\bibitem[{\citenamefont{Wigner}(1973)}]{WignerSIAM25}
\bibinfo{author}{\bibfnamefont{E.~P.} \bibnamefont{Wigner}},
  \bibinfo{journal}{SIAM J. Appl. Math.} \textbf{\bibinfo{volume}{25}},
  \bibinfo{pages}{169} (\bibinfo{year}{1973}).

\bibitem[{\citenamefont{Koster}(1957)}]{KosterPR109}
\bibinfo{author}{\bibfnamefont{G.~F.} \bibnamefont{Koster}},
  \bibinfo{journal}{Phys. Rev.} \textbf{\bibinfo{volume}{109}},
  \bibinfo{pages}{227} (\bibinfo{year}{1957}).

\bibitem[{\citenamefont{Rykhlinskaya and Fritzsche}(2006)}]{RykhlCPC174}
\bibinfo{author}{\bibfnamefont{K.}~\bibnamefont{Rykhlinskaya}}
  \bibnamefont{and}
  \bibinfo{author}{\bibfnamefont{S.}~\bibnamefont{Fritzsche}},
  \bibinfo{journal}{Comput.\ Phys. Comm.} \textbf{\bibinfo{volume}{174}},
  \bibinfo{pages}{903} (\bibinfo{year}{2006}).

\bibitem[{\citenamefont{Brislawn}(1988)}]{BrislawnPAMS104}
\bibinfo{author}{\bibfnamefont{C.}~\bibnamefont{Brislawn}},
  \bibinfo{journal}{Proc. Am. Math. Soc.} \textbf{\bibinfo{volume}{104}},
  \bibinfo{pages}{1181} (\bibinfo{year}{1988}).

\bibitem[{\citenamefont{Gabriel}(1969)}]{GabrielJMP10_1932}
\bibinfo{author}{\bibfnamefont{J.~R.} \bibnamefont{Gabriel}},
  \bibinfo{journal}{J. Math. Phys.} \textbf{\bibinfo{volume}{10}},
  \bibinfo{pages}{1932} (\bibinfo{year}{1969}).

\bibitem[{\citenamefont{Bohanon and Reis}(2006)}]{BohanonJAC23}
\bibinfo{author}{\bibfnamefont{J.~P.} \bibnamefont{Bohanon}} \bibnamefont{and}
  \bibinfo{author}{\bibfnamefont{L.}~\bibnamefont{Reis}}, \bibinfo{journal}{J.
  Algebraic Combin.} \textbf{\bibinfo{volume}{23}}, \bibinfo{pages}{207}
  (\bibinfo{year}{2006}).

\bibitem[{\citenamefont{Dixon}(1970)}]{DixonMC24}
\bibinfo{author}{\bibfnamefont{J.~D.} \bibnamefont{Dixon}},
  \bibinfo{journal}{Math. Comp.} \textbf{\bibinfo{volume}{24}},
  \bibinfo{pages}{707} (\bibinfo{year}{1970}).

\bibitem[{\citenamefont{Babai and R\'onyai}(1990)}]{BabaiMC55}
\bibinfo{author}{\bibfnamefont{L.}~\bibnamefont{Babai}} \bibnamefont{and}
  \bibinfo{author}{\bibfnamefont{L.}~\bibnamefont{R\'onyai}},
  \bibinfo{journal}{Math.\ Comp.} \textbf{\bibinfo{volume}{55}},
  \bibinfo{pages}{705} (\bibinfo{year}{1990}).

\bibitem[{\citenamefont{Zimmerman}(1999)}]{ZimmermGMJ41}
\bibinfo{author}{\bibfnamefont{J.}~\bibnamefont{Zimmerman}},
  \bibinfo{journal}{Glasgow Math. J.} \textbf{\bibinfo{volume}{41}},
  \bibinfo{pages}{115} (\bibinfo{year}{1999}).

\bibitem[{\citenamefont{Dement}(2009)}]{floretion}
\bibinfo{author}{\bibfnamefont{C.~K.} \bibnamefont{Dement}},
  \emph{\bibinfo{title}{Floretions}} (\bibinfo{year}{2009}),
  \urlprefix\url{http://fumba.eu/sitelayout/Floretion.html}.

\bibitem[{\citenamefont{Thyssen et~al.}(2008)\citenamefont{Thyssen, Fleig, and
  Jensen}}]{ThyssenJCP129}
\bibinfo{author}{\bibfnamefont{J.}~\bibnamefont{Thyssen}},
  \bibinfo{author}{\bibfnamefont{T.}~\bibnamefont{Fleig}}, \bibnamefont{and}
  \bibinfo{author}{\bibfnamefont{H.~J.~A.} \bibnamefont{Jensen}},
  \bibinfo{journal}{J. Chem. Phys.} \textbf{\bibinfo{volume}{129}},
  \bibinfo{pages}{034109} (\bibinfo{year}{2008}).

\end{thebibliography}

\end{document}